\documentclass{achemso}
\usepackage[utf8]{inputenc}
\usepackage{graphicx} % Required for inserting images
\usepackage[export]{adjustbox}
\usepackage{indentfirst}
\usepackage{float} 
\usepackage{amsmath,amssymb}
\usepackage{bm}
\usepackage{braket}
\usepackage{color}

\title{Enhancing Accuracy of Quantum-Selected Configuration Interaction Calculations using Multireference Perturbation Theory: Application to Aromatic Molecules}

\author{Soichi Shirai}
\affiliation{Toyota Central Research and Development Laboratories, Incorporated,\\41-1 Yokomichi, Nagakute, Aichi 480-1192, Japan}
\email{shirai@mosk.tytlabs.co.jp}

\author{Shih-Yen Tseng}
\affiliation[QunaSys]
{QunaSys Inc., Aqua Hakusan Building 9F,\\1-13-7 Hakusan, Bunkyo, Tokyo 113-0001, Japan}

\author{Hokuto Iwakiri}
\affiliation[QunaSys]
{QunaSys Inc., Aqua Hakusan Building 9F,\\1-13-7 Hakusan, Bunkyo, Tokyo 113-0001, Japan}

\author{Takahiro Horiba}
\affiliation[ToyotaCRDL]
{Toyota Central Research and Development Laboratories, Incorporated,\\41-1 Yokomichi, Nagakute, Aichi 480-1192, Japan}

\author{Hirotoshi Hirai}
\affiliation[ToyotaCRDL]
{Toyota Central Research and Development Laboratories, Incorporated,\\41-1 Yokomichi, Nagakute, Aichi 480-1192, Japan}

\author{Sho Koh}
\affiliation[QunaSys]
{QunaSys Inc., Aqua Hakusan Building 9F,\\1-13-7 Hakusan, Bunkyo, Tokyo 113-0001, Japan}
\email{koh@qunasys.com}

\begin{document}

\maketitle
\newpage
\section*{Abstract}
Quantum-selected configuration interaction (QSCI) is a novel quantum-classical hybrid algorithm for quantum chemistry calculations. This method identifies electron configurations having large weights for the target state using quantum devices and allows CI calculations to be performed with the selected configurations on classical computers. In principle, the QSCI algorithm can take advantage of the ability to handle large configuration spaces while reducing the negative effects of noise on the calculated values. At present, QSCI calculations are limited by qubit noise during the input state preparation and measurement process, restricting them to small active spaces. These limitations make it difficult to perform calculations with quantitative accuracy. The present study demonstrates a computational scheme based on multireference perturbation theory calculations on a classical computer, using the QSCI wavefunction as a reference. This method was applied to ground and excited state calculations for two typical aromatic molecules, naphthalene and tetracene. The incorporation of the perturbation treatment was found to provide improved accuracy. Extension of the reference space based on the QSCI-selected configurations as a means of further improvement was also investigated.
\newpage
\section{Introduction}
Quantum chemistry calculations play a crucial role in the development of new materials and other scientific advancements. 
Among the various computational methods, density functional theory (DFT) is the most widely used because it provides the balance of tractability and accuracy.~\cite{jones2015density}
However, it is well known that DFT struggles to accurately describe strongly correlated systems, such as complex materials and electronically excited states.
In contrast, wavefunction theory (WFT) offers systematic improvements in accuracy. In the configuration interaction (CI) method, a typical post Hartree-Fock technique based on WFT, the wavefunction is constructed as a linear combination of the Slater determinants corresponding to electron configurations.~\cite{ci1999}
Full CI, which considers all possible configurations arising from combinations of molecular orbitals and electrons, provides an exact solution within the basis set used in the calculations.~\cite{fci1996}
Unfortunately, the number of electron configurations increases factorially with the number of molecular orbitals and electrons, such that full CI is applicable only to very small molecular systems. So-called ``truncated CI" methods that use a limited number of configurations are widely employed to address this issue, although this truncation of the configuration space leads to a loss of accuracy, especially in the case of systems with strong electron correlations.
This dilemma restricts the applications of calculations based on WFT.

Recently, quantum chemical calculations have garnered significant attention as a practical application of quantum computing.~\cite{qcqc2019, qcqc2020-1, qcqc2020-2}
This interest stems from the possibility of preparing quantum superposition states on qubits, potentially enabling calculations equivalent to full CI to be performed in polynomial time. Such advancements are expected to facilitate high-precision quantum chemistry calculations that are otherwise difficult to achieve within the conventional framework. Hence, significant progress in the analysis of complex chemical reactions and the theoretical design of materials could be achieved.
Present-day quantum computers, known as noisy intermediate-scale quantum (NISQ) devices, do not implement error correction for qubits~\cite{nisq2018, nisq2019, nisq2021}, and so calculations performed on NISQ devices are greatly affected by noise, limiting the ability of such calculations to handle deep circuits. On this basis, the variational quantum eigensolver (VQE) algorithm, representing a quantum-classical hybrid combining quantum and classical computing resources, has been proposed.~\cite{Peruzzo:2013bzg, vqe2016, Tilly:2021jem, vqe2022fedorov}
However, there are several challenges that must be overcome before the VQE process can be used for practical applications.
The primary issues associated with this method are statistical fluctuations during the measurement of the expected values together with errors caused by physical noise.
A very large number of samplings are required to suppress the statistical error to a practically acceptable level. In addition, even more sampling is necessary to compensate for the additional statistical errors introduced by error-mitigation techniques used to reduce noise effects.
Error effects can also spoil the variational nature of the VQE algorithm. 
The energy estimated by the quantum device is not guaranteed to provide an upper bound on the exact ground-state energy. As a result, a lower energy value does not necessarily indicate convergence toward the exact ground state.
Other challenges, such as the barren plateau problem that limits optimization of the VQE process, also prevent practical use of this algorithm.

A novel quantum-classical hybrid algorithm referred to as the quantum-selected configuration interaction (QSCI) method has been proposed as a means of addressing these issues.~\cite{Kanno:2023rfr}
A calculation using this method comprises three sequential steps.
In the first step, an input state is generated that provides a rough approximation of the target electronic state.
In the second step, measurements are repeatedly made on this state using a quantum computer to identify electron configurations that contribute significantly to the target state.
Finally, a CI diagonalization calculation using the selected configurations is performed using a classical computer to obtain eigenvalues and eigenvectors.
The QSCI process, when used in conjunction with quantum computers capable of handling quantum superposition states could, in principle, permit the use of large-dimensional configuration spaces as input states. Such spaces are prohibitively difficult to manage with classical computers.
Because the CI diagonalization calculation in the final step is performed using a classical computer, the resulting energy is not affected by noise originating from the quantum circuit, and so always represents an upper limit for the exact ground state energy.
That is, the QSCI method can employ large configuration spaces while suppressing errors in the calculated values originating from noise, the latter of which is a major drawback of calculations using an NISQ device.
The QSCI technique can be applied to calculations of both ground and excited states by either expanding the subspace or repeating the procedure for each eigenstate.

Even though the QSCI algorithm can mitigate some of the problems associated with VQE calculations, qubit noise remains a problem. This noise occurs during the preparation of the input state and during measurements, limiting the use of this process to only a small configuration space.
Hence, it is almost impossible to incorporate dynamical electron correlations into this process, meaning that QSCI alone cannot be used to carry out such calculations with quantitative accuracy.
To overcome this drawback, we have proposed a computational approach to improve accuracy based on performing multireference theoretical calculations on a classical computer using a configuration space selected by the QSCI method as its reference.
The present study applies this methodology to calculations involving multireference perturbation theory (MRPT) with moderate computational costs. 
This technique, referred to herein as the QSCI-PT method, combines the well-established quasi-degenerate perturbation theory with general multi-configuration reference functions (GMC-QDPT)~\cite{nakano2002} with the QSCI process.
Several prior studies have investigated the integration of perturbation theory with the VQE framework and a number of papers on this topic have been published~\cite{PhysRevLett.129.120505, ryabinkin2021posteriori, tammaro2023n, li2023toward, liu2024perturbative, di2024platinum}.
In the QSCI-PT method, electron configurations having large weights are first selected through the QSCI sampling process.
Following this, a QSCI-referenced MRPT treatment is used to incorporate dynamical electron correlations.
Using this approach, the GMC-QDPT method is likely the most suitable technique for combination with the QSCI algorithm for several reasons.
Because the QSCI space consists of selected configurations, it is not possible to specify in advance the structure of the configuration space.
In this regard, with the GMC-QDPT method, MRPT calculations can be performed using any type of configuration space.
Another important advantage of adopting the GMC-QDPT method is that the configuration space initially prepared using the QSCI technique can be extended via augmentation with additional electron configurations.
This modification can compensate for the shortcomings of the current QSCI method when used in conjunction with NISQ devices that allow only a small configuration space.
The complementary electron configurations added to the QSCI space can be obtained systematically using configurations in the QSCI space as parent configurations.

\begin{figure}[H]
    \begin{center}
        \includegraphics[width=10cm]{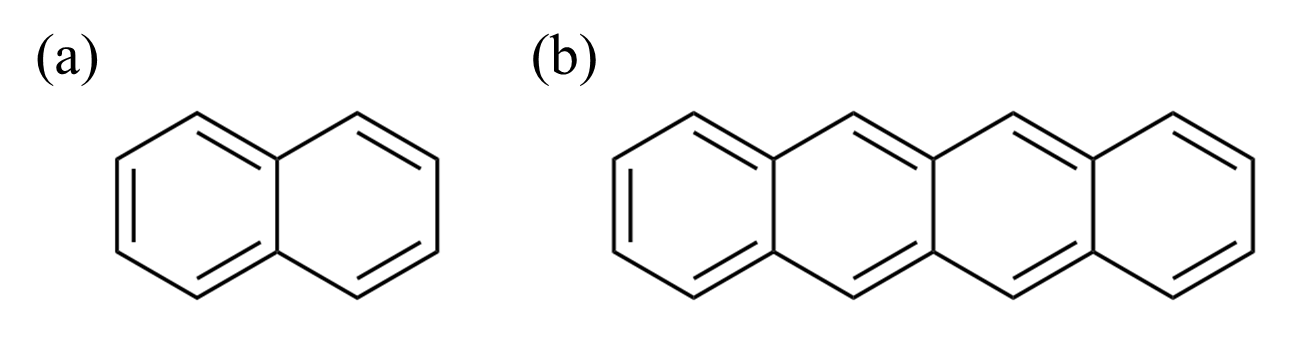}
        \caption{Molecules for which calculations were performed in this study: (a) naphthalene and (b) tetracene.}
        \label{fig:naph-tetra}
    \end{center}
\end{figure}

The present work applied the QSCI-PT technique to the analysis of the excited states of two typical aromatic molecules: naphthalene and tetracene (Figure \ref{fig:naph-tetra}). The accuracy of the computational results was found to be significantly improved compared with the accuracy provided by the original QSCI method.
Aromatic molecules are the building blocks of the organic materials commonly used in optical and electronic devices such as light-emitting diodes,~\cite{el1987, el1993, el1999} solar cells,~\cite{solar-cell-2007, solar-cell-2010-1, solar-cell-2010-2} and semiconductors~\cite{hole-transport-2007, electron-trasnport-2010}. Hence, a detailed understanding of the ground and excited states of these compounds is important not only for basic science but also for the development of new functional materials.
Because the electronic states of aromatic molecules having $\pi$-conjugated electron systems generally exhibit significant multiconfiguration character, these molecules were considered suitable model compounds to evaluate the performance of the QSCI-PT process.

\section{Theory}
\subsection{The quantum-selected configuration interaction method}\label{sec:2_1}
The hybrid quantum-classical QSCI algorithm can be used to calculate electronic states on noisy quantum devices.
The goal of this process is to select primary electron configurations from a set of input states prepared on quantum devices to construct the desired electronic states via diagonalization of the CI matrix with a classical computer.
Here, we use the construction of an electronic ground state of the second quantized Hamiltonian $\hat{H}$ as a simple example.~\cite{fock:1932} In this process, an input state $\ket{\psi^{0}_{\mathrm{in}}}$ approximating the electronic ground state is initially generated using, as an example, the VQE algorithm~\cite{Peruzzo:2013bzg}.
In the case that $\ket{\psi^{0}_{\mathrm{in}}}$ is prepared using a quantum circuit with $N_q$ qubits based on the Jordan-Wigner transformation~\cite{Jordan1928berDP, Nielsen2005TheFC}, one then measures it in the computational basis and obtains the output as a bit string corresponding to the electron configurations. 
Repeating the sampling procedure $N_{\mathrm{shot}}$ times with a quantum computer produces a set of electron configurations, each occurring with a different frequency. Based on the sampling results, the set $\mathcal{S}_{R}$ is defined, comprising the $R$ most  frequently observed electron configurations.
The effective Hamiltonian, $\bm{H}_R$, is subsequently constructed in the subspace spanned by $\mathcal{S}_{R}$, with the elements calculated as
\begin{align}\label{eq:H_eff}
 (\bm{H}_R)_{xy}= \braket{x|\hat{H}|y}~\mathrm{ for}~\ket{x}, \ket{y} \in \mathcal{S}_R.
\end{align}
The eigenvalue equation is then solved as
\begin{align}
    \bm{H}_R\bm{c} = E_R\bm{c},
\end{align}
where $\bm{c}$ is the eigenvector having eigenvalue $E_R$ that satisfies $\bm{c}^{\dag}\bm{c} = 1$. Here, $E_R$ and $\bm{c}$ are approximations of the exact ground state energy and the corresponding CI coefficients, respectively.
The final output state, $\ket{\psi^{0}_{\mathrm{out}}}$, is then constructed as
\begin{align}
    \ket{\psi^{(0)}_{\mathrm{out}}} = \sum_{\ket{x}\in \mathcal{S}_R} c_x \ket{x},
    \label{eq:output_ground}
\end{align}
where $c_x$ are the elements of the eigenvector $\bm{c}$ for the corresponding electron configurations and $\ket{\psi^{(0)}_{\mathrm{out}}}$ is an approximation of the exact ground state of $\hat{H}$.

\begin{figure}[H]
\includegraphics[width=0.9\textwidth,center]{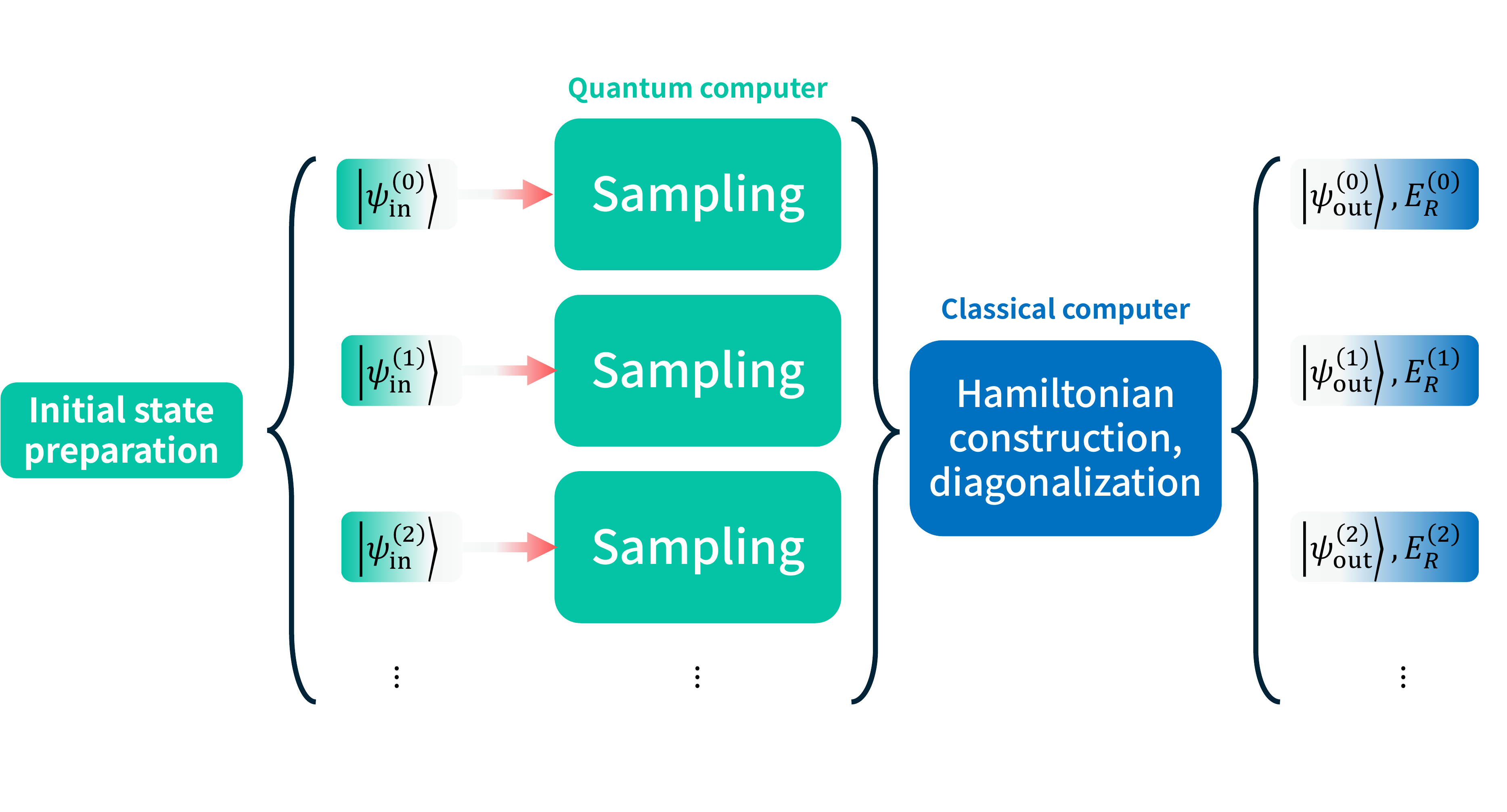}
    \caption{A schematic summarizing the QSCI calculation process.}
    \label{fig:analysis_procedure}
\end{figure}

Here, we examine a means of extending the procedure described above to excited states. 
This work employed the so-called ``single diagonalization scheme" previously described by Kanno {\it et al.}~\cite{Kanno:2023rfr} to obtain the electronic excited states in conjunction with the QSCI approach (Figure \ref{fig:analysis_procedure}).
In this process, the capture of electronic excited states required additional input states for constructing a common subspace comprising ground and excited states. In the case that the intent is to search for the $N_s$ lowest energy eigenstates of a Hamiltonian $\hat{H}$, the input states $\ket{\psi^{(k)}_{\mathrm{in}}}$ $(k=0,1,\cdots,N_s - 1)$ have to be prepared to  approximate the true eigenstates. Following this, sampling is carried out to obtain a set, $S^{(k)}_{R_k}$, consisting of the $R_k$ most important configurations for the $k$-th input state. Finally, these configuration sets are combined to form the common subspace
\begin{align}\label{eq:common_subspace}
    \mathcal{S}_{R} = \mathcal{S}^{(0)}_{R_0} \cup \mathcal{S}^{(1)}_{R_1} \cdots \cup \mathcal{S}^{(N_s - 1)}_{R_{N_s - 1}},~ R = \sum^{N_s-1}_{k=0} R_{k}.
\end{align}
After obtaining the selected basis states, the $R\times R$ Hermitian matrix $\bm{H}_R$ is generated using Eq.\,\eqref{eq:H_eff}. Following this, the corresponding energy eigenvalues $E^{(k)}_{R}$ and eigenstates $\bm{c}^{(k)}_{R}$ with $k = 0,\cdots, N_s-1$ are acquired after performing the diagonalization of $\bm{H}_R$. The output states are then constructed as
\begin{align}
    \ket{\psi^{(k)}_{\mathrm{out}}} = \sum_{\ket{x}\in \mathcal{S}_R} c^{(k)}_x \ket{x},~k=0,\cdots,N_s-1.
    \label{eq:output_states}
\end{align}

It should be noted that a post-selection process for sampled electron configurations can be employed to suppress device noise. Symmetries in the structure of the electronic Hamiltonian correspond to conserved quantities, such as the total electron number, $N_e$, and the $z$ component of the total spin, $S_z$. 
During sampling with an actual quantum device, physical noise sources such as the bit-flip noise and readout error can potentially contaminate the selected configurations, thereby spoiling the symmetry sector $(N_e,S_z)$ of the target state.
Fortunately, post-selections of the sampling outcomes can be performed simply by discarding configurations that do not have the desired $(N_e,S_z)$ values.

\subsection{GMC-QDPT combined with the QSCI method}\label{sec:2_2}

As noted in the Introduction, full CI calculations that consider all electron configurations arising from combinations of molecular orbitals and electrons will have an excessive computational cost for large systems.
Therefore, for reasons of practicality, the methods using limited electron configurations has been widely adopted.
One of the most typical approaches to extract configurations from the full CI space is to define the molecular orbitals that are essential in describing the electronic state of interest as active orbitals and the electrons in these orbitals as active electrons. The electron configurations that are generated by distributing these active electrons over the active orbitals are then used.
The configuration space consisting of all electron configurations arising from combinations of active orbitals and electrons is referred to as the complete active space (CAS) while the CI method using the CAS is denoted as the CASCI.~\cite{casci2021} This is a typical multiconfiguration theory.
The complete active space self-consistent field (CASSCF) method, which combines the CASCI with the optimization of molecular orbitals, is also widely used.~\cite{casscf1980}
However, since the QSCI space generally involves only selected configurations, CI calculations on a classical computer should be subsequently performed using an incomplete active space.
In order to enable CI and MRPT calculations based on an arbitrarily structured configuration space, the present work employs the QSCI algorithm together with the GMC-QDPT method.
The GMC-QDPT method is described to be a multireference perturbation theory based on a general configuration space including the CAS.

As discussed, present-day NISQ devices cannot handle a large number of active orbitals, as this requires a significant quantity of qubits.
In addition, in the case of calculations using actual quantum devices, large contributing configurations may not be selected as a result of noise effects.
Consequently, both dynamic and static electron correlations may not be considered to a sufficient extent.
For these reasons, as a practical measure, we propose to augment the configuration space by incorporating complementary electron configurations generated based on the configurations selected by the QSCI method.
The configurations selected by QSCI are set as the parent configurations while some occupied and unoccupied orbitals are also selected as active orbitals.
Complementary configurations are subsequently generated by electron excitations from the parent configurations.
Following this, GMC-QDPT calculations based on the augmented configuration space and involving both the parent and complementary configurations are carried out.
It should be noted that the original GMC-QDPT process assumes that the wave function optimized for molecular orbitals using the MCSCF method is adopted as a reference.
In contrast, this study uses the CI wave function as a reference and does not optimize the molecular orbitals for the target electronic states. Instead, molecular orbitals obtained using the Hartree-Fock method are employed.
The Computational detail section describes the calculation procedures and conditions in greater detail.

\subsection{Excited states of naphthalene and tetracene}\label{sec:2_3}
The low-lying singlet excited states of naphthalene are characterized by two excited states, written as $^{1}$L$_{\rm{a}}$ and $^{1}$L$_{\rm{b}}$ in Platt's notation.~\cite{platt1949}
The four orbitals going from the second highest occupied molecular orbital (HOMO--1) to the second lowest unoccupied molecular orbital (LUMO+1) are relevant to the main configurations of these excited states.~\cite{roos1994, hirao1996}

\begin{figure}[H]
    \begin{center}
        \includegraphics[width=11cm]{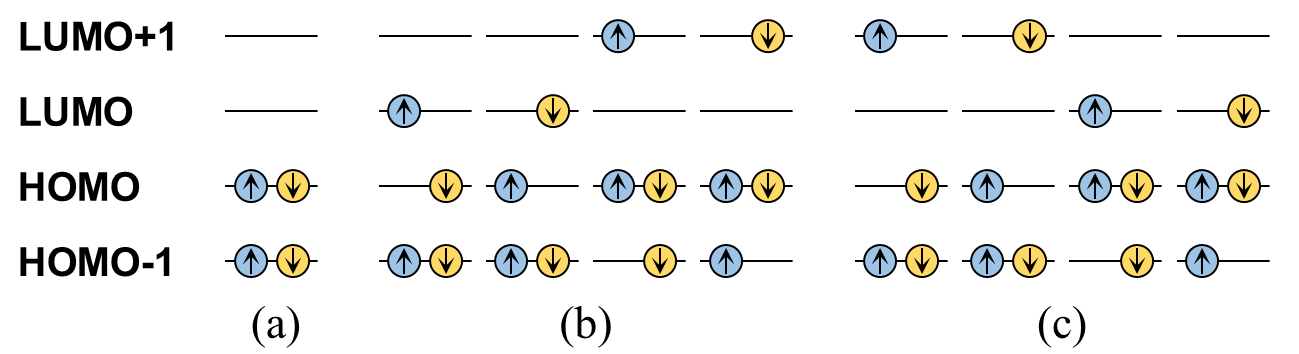}
        \caption{Primary configurations for the (a) ground state, (b) $^{1}\rm{L_{a}}$ excited state, and (c) $^{1}\rm{L_{b}}$ excited state. Here, HOMO is the highest occupied molecular orbital and LUMO is the lowest unoccupied molecular orbital while HOMO--1 is the second HOMO and LUMO+1 is the second LUMO.}        \label{fig:La-Lb-main-config}
    \end{center}
\end{figure}

The main configurations of the $^{1}$L$_{\rm{a}}$ state are the HOMO $\rightarrow$ LUMO and HOMO--1 $\rightarrow$ LUMO+1 singly excited configurations (Figure \ref{fig:La-Lb-main-config}), with the former making a greater contribution.
In contrast, the main configurations of the $^{1}$L$_{\rm{b}}$ state are the HOMO $\rightarrow$ LUMO+1 and the HOMO--1 $\rightarrow$ LUMO singly excitation configurations, both of which contribute almost equally.
The energy level of the $^{1}$L$_{\rm{a}}$ state is also higher than that of $^{1}$L$_{\rm{b}}$ by approximately 0.5 eV.~\cite{grimme2003}

Employing the QSCI method requires clarification of the conditions (such as the initial state and ansatz) under which the configurations with the most suitable large contributions (including the main configuration) are selected for the target electronic states.
The selection of the main configurations for the $^{1}$L$_{\rm{a}}$ and $^{1}$L$_{\rm{b}}$ states of naphthalene is an important aspect of the investigation of conditions.

The transferability of the resulting computational scheme is also examined by calculating the excited states of tetracene in the same manner.
The $^{1}\rm{L_{a}}$ and $^{1}\rm{L_{b}}$ excited states of tetracene have the same main configurations as the $^{1}\rm{L_{a}}$ and $^{1}\rm{L_{b}}$ excited states of naphthalene, respectively.~\cite{kawashima1999}
In contrast, the energetic ordering for the $^{1}\rm{L_{a}}$ and $^{1}\rm{L_{b}}$ states of tetracene is the inverse of that of naphthalene.~\cite{grimme2003}

\section{Computational details}

The molecular structure of naphthalene was optimized by applying $D_{2h}$ symmetry and using the B3LYP~\cite{b3lyp1980, b3lyp1988, b3lyp1993, b3lyp1994} functional and the 6-31G(d) basis set~\cite{631g1971, 631g1972, 631gd}.
This level of calculation allows the molecular structure of naphthalene to be optimized with a high degree of accuracy.~\cite{shiraiyanai2016}
The molecular geometry of an isolated naphthalene molecule was previously ascertained using a molecular beam technique~\cite{nph-geom} and the results of the present calculations agreed with the experimental values for bond lengths within 0.5\% and for bond angles within 0.2\%.
The molecular structure of tetracene was also optimized at the same level of theory and the results were compared with crystallographic data.~\cite{tetra-bonds, tetra-angles}
The bond lengths and bond angles agreed with the experimental values within 2.3\% and 1.3\%, respectively.
The geometry optimization calculations were performed using the Gaussian16 program~\cite{g16} and calculations employing the cc-pVDZ basis set~\cite{dunning1989} were subsequently carried out for these molecular structures.
The active space for the QSCI calculations was constructed by specifying the HOMO--1, HOMO, LUMO and LUMO+1 as the active orbitals.
These four orbitals are denoted herein as $2\pi, 1\pi, 1\pi^{*}$, and $2\pi^{*}$, respectively.
The electronic ground state as well as the $^{1}$L$_{\rm{a}}$ and $^{1}$L$_{\rm{b}}$ states were calculated for both naphthalene and tetracene.

This work utilized the open source library QURI Parts~\cite{quri2024} to implement all the necessary codes required to execute the QSCI algorithm and qiskit~\cite{qiskit2024} to run \texttt{ibmq\_qasm\_simulator} and \texttt{ibm\_osaka}.
The GMC-QDPT calculations were carried out using the GAMESS program.~\cite{gamess1993, gamess2005, gamess2020}

\subsection{Preparation of input states for QSCI calculations}\label{sec:3_1}
To focus on the performance of the QSCI-PT algorithm in this study, we adopted noiseless simulation results obtained from the variational quantum deflation (VQD) algorithm~\cite{Higgott:2018doo} to prepare the input states. 
The VQD cost function was expressed as
\begin{align}
L(\bm{\theta}_k)=\braket{\psi(\bm{\theta}_k)|H|\psi(\bm{\theta}_k)} + \sum^{k-1}_{j=0}\beta_{j}|\braket{\psi(\bm{\theta}_k)|\psi(\bm{\theta}_j^{\mathrm{opt}})}|^2,
\label{eq:vqd_cost}
\end{align}
where $H$ is a modified Hamiltonian combining the molecular Hamiltonian, $H_{\mathrm{mol}}$, and a penalty term for the total spin, $\hat{S}^2$. This is written as
\begin{align}
H = H_{\mathrm{mol}} + \alpha\hat{S}^2.
\end{align}
The coefficient for the penalty term was set to $\alpha = 3.0$, ensuring that all optimized states were singlet states. The second term in Eq.~\ref{eq:vqd_cost} was a penalty for the $k$-th excited state $\ket{\psi(\bm{\theta}_k)}$. This term penalized overlap with the optimized ground state and all excited states $\ket{\psi(\bm{\theta}_j^{\mathrm{opt}})}$ below $\ket{\psi(\bm{\theta}_k)}$, ensuring orthogonality among the states. When using an expressive ansatz, it is generally sufficient to select a value for the weight in the penalty term of $\beta_{j} > E_{k}-E_{i}$, which guarantees a minimum at $E_{k}$~\cite{Higgott:2018doo, Kuroiwa2021, Shirai2022}. In the work reported herein, a value of $\beta_j=3.0$ was used for all $j$ to ensure that the minimum of the cost function corresponded to the energy of the $k$-th excited state.

\begin{figure}[H]
\centering

\includegraphics[width=0.6\textwidth,center]{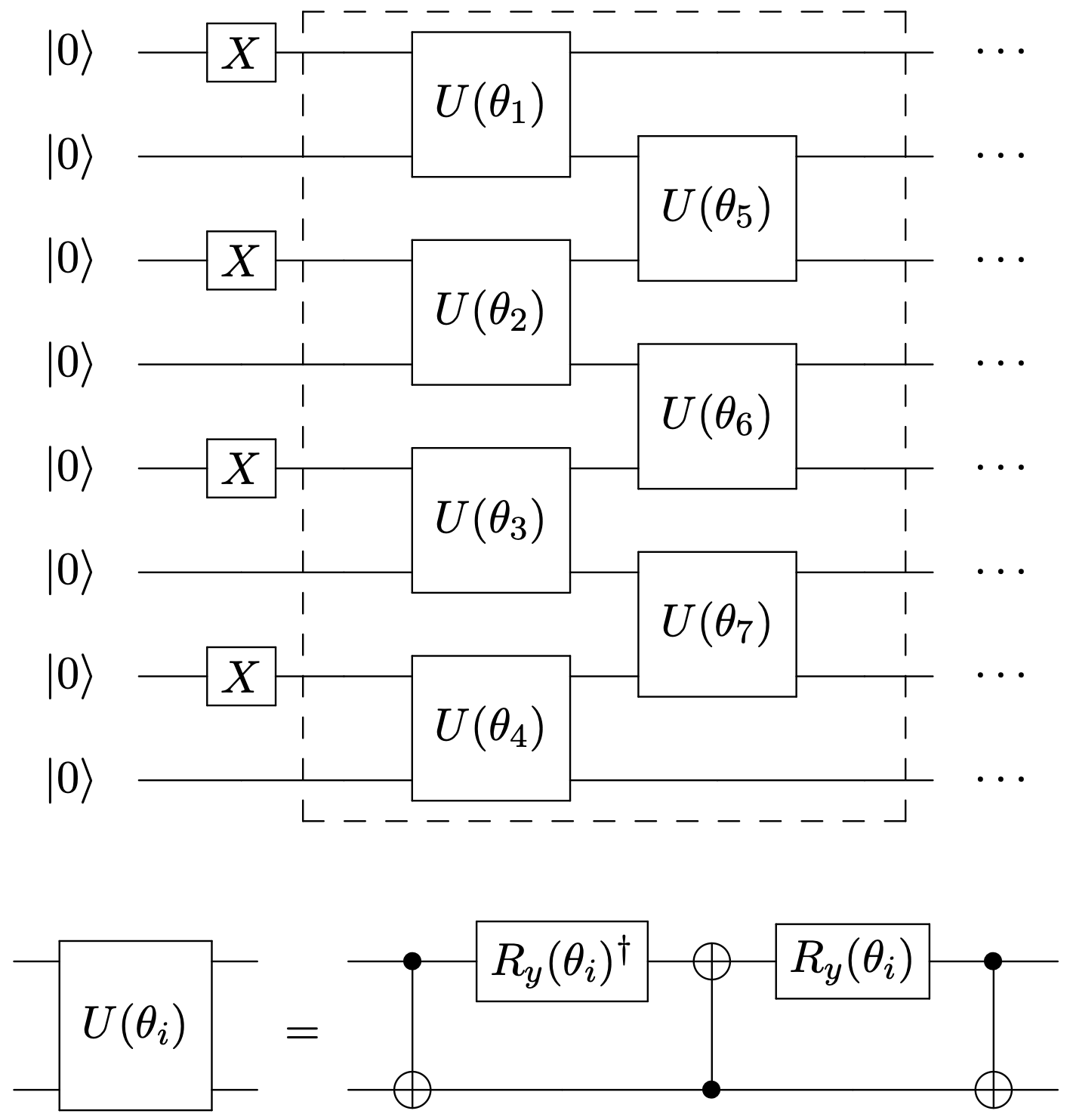}

\caption{A schematic plot showing the eight qubit real-valued symmetry-preserving (RSP) ansatz used in the preparation of initial states. The dashed box including the seven $U$ gates is repeated $d$ times to generate an RSP ansatz of depth $d$. The depth was set to $d=3$ in this work. The set of parameters $\bm{\theta}=(\theta_1, \theta_2, \theta_3,\cdots)$ was optimized. The $U(\theta_{i})$ gates $(i = 1, 2, 3, \cdots)$ in the RSP ansatz comprise three CNOT gates interleaved with two $R_y$ gates.}
\label{fig:U_gate}

\label{fig:ansatz_SPR}
\end{figure}

The real-valued symmetry-preserving (RSP) ansatz~\cite{gard2020efficient,ibe2022} was used for the parametrized quantum states $\ket{\psi(\bm{\theta_k})}$. The RSP ansatz has the advantage of conserving the electron number, $N_e$, the $z$ component of the total spin, $S_z$, and the time-reversal symmetry~\cite{gard2020efficient}. During these calculations, the circuits could be represented using eight qubits because four molecular orbitals were employed for the active space in conjunction with the Jordan-Wigner transformation. The RSP ansatz was composed of $X$ gates for those qubits representing the occupied spin orbitals and a set of $U$ gates, repeated $d$ times for a depth-$d$ RSP ansatz (Figure~\ref{fig:ansatz_SPR}). A viable circuit depth allowing the ansatz to be executed on the actual quantum device \texttt{ibm\_osaka} was determined by producing a rough estimation of the fidelity of the circuit.
This estimation assumed that the fidelity of the two qubit gates was the most important factor for determining the overall fidelity. Consequently, the total fidelity was expressed as
\begin{align}
    f = \left(\text{2-qubit~gate~fidelity}\right)^{3n_U\cdot d},
\end{align}
where $n_U$ is the number of $U$ gates per depth, equal to seven in the present work, and $d$ is the number of depths.
Using the median two qubit ECR gate fidelity value for the \texttt{ibmq\_osaka} calculations (approximately 99.0456$\%$ as of January 26, 2024) it was estimated that the circuit depth had to be $d \lesssim 3$ for the total fidelity to exceed $f=50\%$. 
We found that the depth-$1$ and depth-$2$ RSP ansatzes were too shallow and lacked sufficient expressibility for the target states. 
Therefore, to account for the fidelity limitations of the two qubit gates, the depth was set to $d=3$ in the subsequent analysis.
The initial values of the parameter set $\bm{\theta}$ were chosen at random from within the range $[0,4\pi]$ during the optimization procedure and the cost function was optimized using the ``minimize" function in the ``scipy" library.~\cite{2020SciPy-NMeth} The Broyden-Fletcher-Goldfarb-Shanno (BFGS) algorithm~\cite{Broyden1970TheCO, Fletcher1970, Goldfarb1970, Shanno1970, bfgs2019} was employed as the optimizer. The VQD technique was applied to the naphthalene and tetracene molecules and the RSP ansatz for each was optimized to provide input states for the QSCI process.

\subsection{QSCI and QSCI-PT calculations}\label{sec:3_2}

The VQD algorithm was used to prepare the input states as the approximations of the electronic ground and excited states.
Subsequently, the QSCI algorithm was applied to these states to construct the effective Hamiltonian $\bm{H}_R$, as described in the Theory section, on both the \texttt{ibmq\_qasm\_simulator} simulator and the actual quantum device \texttt{ibm\_osaka}. The total number of shots was set to 9,999 and was divided equally among the ground state and the two excited states when sampling electron configurations.

QSCI calculations were performed by diagonalizing $\bm{H}_R$, from which approximations of the exact energies and their corresponding eigenstates were extracted. 
The QSCI-PT calculations were performed using the electron configurations selected with \texttt{ibm\_osaka}.
In the following text, the calculations based on this configuration space are referred to as the QSCI and QSCI-PT calculations, unless otherwise noted.

QSCI-PT calculations were initially performed for a naphthalene molecule employing a QSCI reference space.
GMC-QDPT calculations with CISD(4e, 4o) and CAS(4e, 4o) reference spaces were also performed for comparison.
CISD(4e, 4o) represented the configuration space involving the closed-shell singlet Hartree-Fock configuration and the excited configurations resulting from single and double excitations from the Hartree-Fock state within the four active orbitals.
The dimension of CISD(4e, 4o), meaning the number of Slater determinants, was 27 in the case that the symmetry of electronic states was not taken into account.
CAS(4e, 4o) comprised all the electron configurations generated from combinations of the four active orbitals and four active electrons and therefore had a dimension of 36.

\begin{figure}[H]
    \begin{center}
        \includegraphics[width=8.25cm]{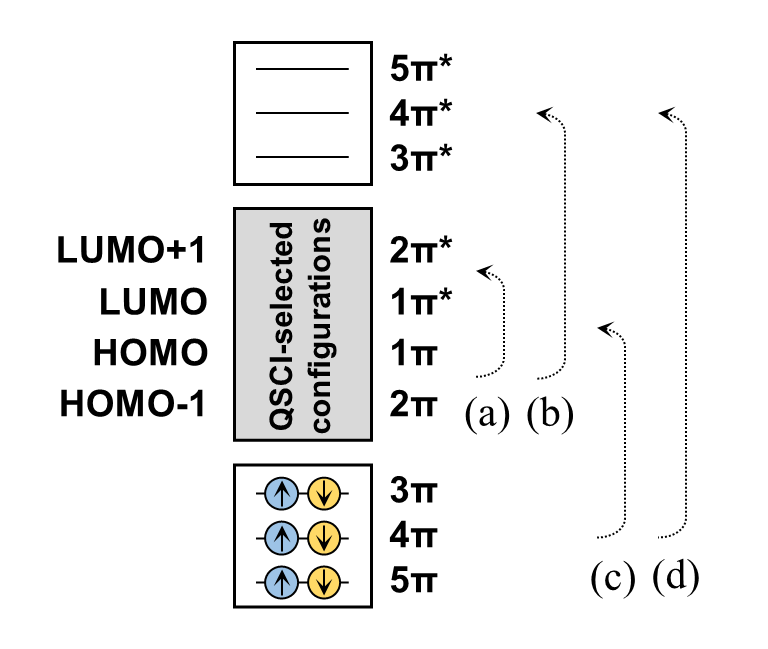}
        \caption{Construction of QSCI-selected configurations + X reference spaces for QSCI-PT calculations of a naphthalene molecule. HOMO--1, HOMO, LUMO, and LUMO+1 correspond to 2$\pi$, 1$\pi$, 1$\pi$*, and 2$\pi$* orbitals, respectively. Shown are excitations within orbitals from (a) 2$\pi$ to 2$\pi$*, (b) 2$\pi$--2$\pi$* to 3$\pi$*--5$\pi$*, (c) 3$\pi$--5$\pi$ to 2$\pi$--2$\pi$*, and (d) 3$\pi$--5$\pi$ to 3$\pi$*--5$\pi$*. The total excitation number is the sum of the excitation numbers for each of these. The upper limit for the total excitation number was set to two for X = SD, three for X = SDT, and four for X = SDTQ.}        \label{fig:qscipt-ref}
    \end{center}
\end{figure}

The augmented reference spaces for naphthalene were prepared by first setting the configurations involved in the QSCI process as the parent configurations.
Following this, three $\pi$ orbitals (5$\pi$, 4$\pi$, and 3$\pi$ in order of increasing energy) and three $\pi^{*}$ orbitals (3$\pi^{*}$, 4$\pi^{*}$, and 5$\pi^{*}$ in order of increasing energy) were also selected as the active orbitals to give a total of 10 active orbitals.
Accordingly, six electrons in these three $\pi$ orbitals were additionally set as active electrons and therefore the total number of active electrons was increased to 10.
Complementary configurations were subsequently generated via the excitation of electrons from the parent configurations.
The GMC-QDPT calculations were then carried out using a configuration space involving both the parent and complementary configurations and the augmented reference spaces are denoted herein as QSCI + X.
As depicted in Figure \ref{fig:qscipt-ref}, this process involved four types of excitation within the original active orbitals from 2$\pi$ to 2$\pi$*, from 2$\pi$--2$\pi$* to 3$\pi$*--5$\pi$*, from 3$\pi$--5$\pi$ to 2$\pi$--2$\pi$*, and from 3$\pi$--5$\pi$ to 3$\pi$*--5$\pi$*. The total excitation number was the sum of the excitation numbers for these four types.
The upper limit for the total excitation number from the parent configuration was set to two for X = SD, three for X = SDT, and four for X = SDTQ.
GMC-QDPT calculations with CAS(10e, 10o) reference spaces based on the same active orbitals and electrons were also performed for comparison.
In the perturbation calculations, electrons in the molecular orbitals derived from C 1\textit{s} were excluded from the correlations.
The intruder state avoidance (ISA) technique~\cite{isa-witek} was applied and the shift value was set to 0.02.

Calculations for a tetracene molecule were performed in the same manner except for the active orbitals and active electrons associated with the augmented reference space.
The complementary configurations were generated by selecting seven $\pi$ orbitals and seven $\pi$* orbitals other than the four orbitals from HOMO--1 to LUMO+1 as additional active orbitals. The 14 electrons in these additional $\pi$ orbitals were newly set as active electrons.
Therefore, the augmented reference space for tetracene was derived from 18 active orbitals and 18 active electrons.
Calculations involving the CAS(18e, 18o) reference space were not performed because the dimensions of this space were prohibitively large. Henceforth, the notation for a reference space may also denote CI calculations using it. 

\section{Results and discussion}
\subsection{Configuration selection using a quantum computer}
\begin{figure}[H]
\centering
\includegraphics[width=6.0in]{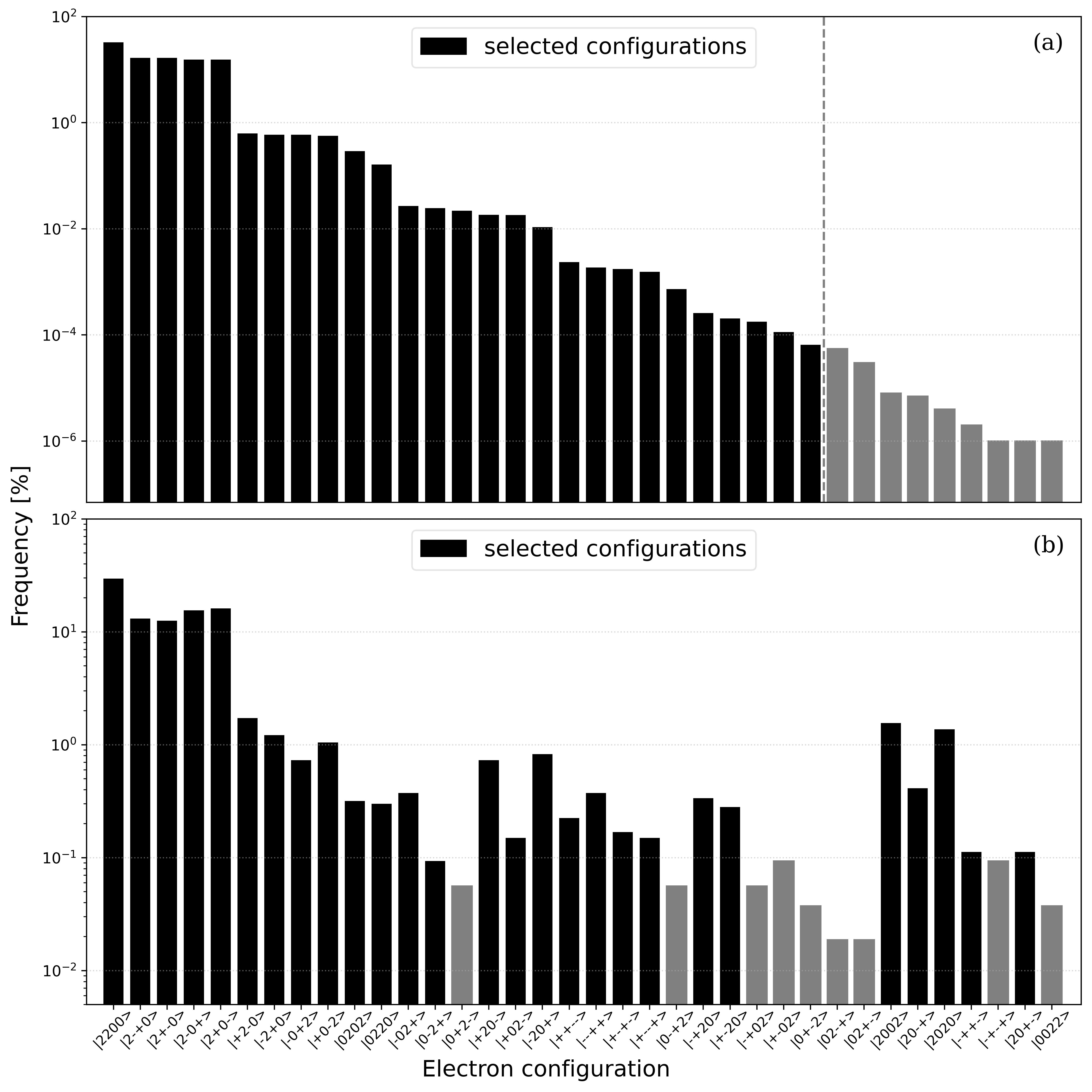}
\caption{The appearance frequencies of electron configurations for the naphthalene molecule as obtained from sampling using the (a) noiseless stimulation and (b) actual quantum device \texttt{ibmq\_osaka} following post-selection of the total electron number, $N_e$, and the $z$-component of the total spin, $S_z$.
The vertical dashed line in (a) separates the 27 selected configurations from the unused ones. The electron configuration labels along the $x$-axis are arranged in descending order based on the magnitude of their coefficients as obtained from the noiseless simulation.}
\label{fig:combined_sampling_result}
\end{figure}

This section presents the results of the sampling of input states prepared using the VQD method. Note that a detailed discussion of the VQD setup is provided in the Computational details section.
The sampling results for a naphthalene molecule from the noiseless simulator and the \texttt{ibmq\_osaka} are presented in Figure~\ref{fig:combined_sampling_result}. 
In the case of the noiseless simulation, the electron number, $N_e = 4$, and the \textit{z} component of electron spin, $S_z = 0$, were guaranteed for every configuration because the SPR ansatz, which conserves $N_e$ and $S_z$, was used for these calculations.
In contrast, when employing the \texttt{ibmq\_osaka} approach, the electron configurations that did not meet these conditions appeared in the sampling results even though the SPR ansatz was applied because of the noise in the NISQ device. 
To address this issue, undesired configurations introduced by noise associated with the operation of the quantum computer were removed through post-selections so as to eliminate all configurations without the correct $N_e$ and $S_z$. 
As noted in the Theory section, an effective Hamiltonian, $\bm{H}_R$, was constructed from the sampling results and diagonalized using a classical computer. The quantity of the most important configurations, $R = 27$, was chosen as the dimension of $\bm{H}_R$ and was equal to the number of CISD(4e, 4o) configurations. In this figure, each plot represents the combined electron configuration sampling of the ground state and two excited states, with selected configurations shown in black and removed configurations in gray. The $x$-axis labels are the electron configurations, in the same order as shown in Figure ~\ref{fig:combined_sampling_result}(a) for the noiseless simulation.
Note also that a simplified notation is used herein to describe the electron configurations.
Based on ket notation, the doubly-occupied, $\alpha$-singly-occupied, $\beta$-singly-occupied, and vacant orbitals are denoted by 2, +, --, and 0, respectively.
As an example, $\ket{2+-0}$ indicates an electron configuration with two electrons in the HOMO--1, one electron with $\alpha$-spin in the HOMO, one electron with $\beta$-spin in the LUMO, and no electron in the LUMO+1.
Both the results from the noiseless simulator and the real quantum device show that all the main configurations presented in Figure \ref{fig:La-Lb-main-config} were successfully selected through the sampling.
Notably, despite being affected by noise, the electron configurations with large weights for the target electronic states, such as the main configurations, could also be selected by the real device when employing the ansatz and computational conditions adopted here.
This outcome indicates that an essential step toward the practical application of the QSCI algorithm using actual quantum devices, the selection of main configurations, has been successfully addressed.
It is of interest that some doubly-excited configurations were not selected whereas triply-excited configurations, which were not included in the CISD(4e, 4o) configuration space, were chosen.
The results from the simulator suggest that this was not the result of noise.
In the QSCI approach, electronic configurations making significant contributions to the target electronic state can, in principle, be selected regardless of the number of excited electrons if we can prepare the input states including these electronic configurations.
As a result, this method may allow more efficient incorporation of the effects of multiply-excited configurations compared with approaches that specify the number of excited electrons.

The difference in selected configurations between the noiseless simulation and the \texttt{ibmq\_osaka} sampling is evident in Figure ~\ref{fig:combined_sampling_result}.
For the \texttt{ibmq\_osaka} sampling results in Figure~\ref{fig:combined_sampling_result}(b), whose original configuration distribution was contaminated by noise-induced configurations prior to post-selection, exhibit behavior that is distinct from the noiseless sampling data shown in Figure~\ref{fig:combined_sampling_result}(a).
All eight triply-excited configurations were selected by the noiseless sampling. 
In contrast, three of eight triply-excited configurations and two doubly-excited configurations were not chosen by the actual quantum device. Instead, an additional five doubly-excited configurations were selected, representing an issue unique to configuration selection by NISQ devices.
Such noise-induced substitutions are more likely to occur in electronic configurations with relatively small contributions. 
Although the resulting loss of accuracy is a potential problem, the affected configurations make relatively small contributions in this eight qubit system and so the overall impact may be negligible.
Even so, as the number of qubits increases, noise effects will become more significant.
One possible solution is to compensate for this effect by expanding the configuration space on a classical computer and this approach is demonstrated in the next section.

\begin{figure}[H]
\centering
\includegraphics[width=6.0in]{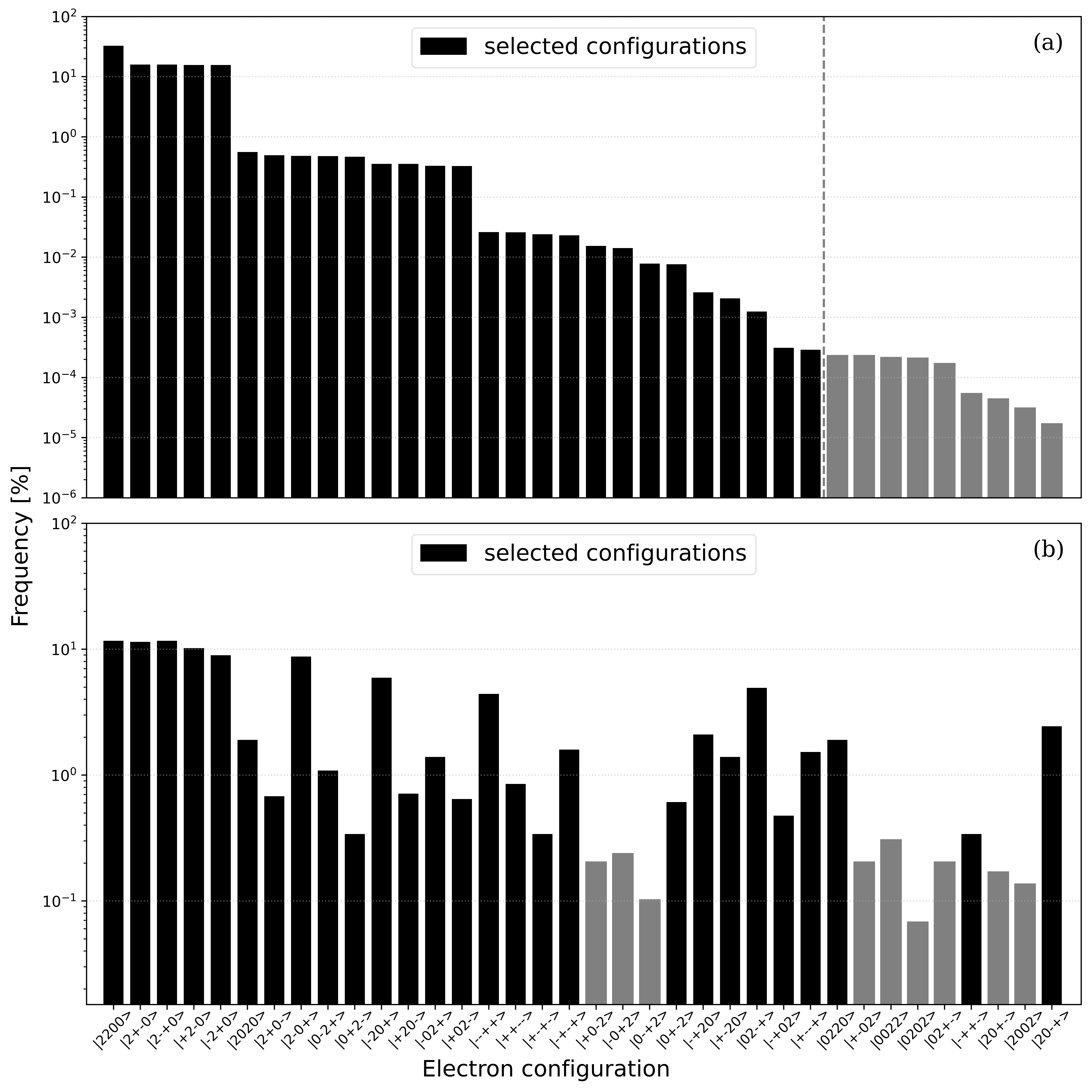}
\caption{The appearance frequencies of electron configurations for the tetracene molecule as obtained from sampling using the (a) noiseless stimulation and (b) actual quantum device \texttt{ibmq\_osaka}, following post-selection of the total electron number, $N_e$, and the $z$-component of the total spin, $S_z$.
The vertical dashed line in (a) separates the 27 selected configurations from the unused ones. The electron configuration labels along the $x$-axis are arranged in descending order based on the magnitude of their coefficients as obtained from the noiseless simulation.} \label{fig:combined_sampling_result_tetracene}
\end{figure}

The samplings for a tetracene molecule were carried out based on the same scheme that was applied to naphthalene and the results for the noiseless sampling and that employing \texttt{ibmq\_osaka} are presented in Figure ~\ref{fig:combined_sampling_result_tetracene}.
The resulting behaviors were similar to those observed in the trial with naphthalene (Figure ~\ref{fig:combined_sampling_result}).
The Hartree-Fock type ground configuration and all eight one-electron excitation configurations were selected.
Thus, the main configurations for the three electronic states of interest were all involved in the QSCI configuration space.
In the noiseless sampling results, ten doubly-excited configurations and eight triply-excited configurations were selected.
In contrast, the \texttt{ibmq\_osaka} sampling did not select three of the eight triply-excited configurations. Rather, doubly-excited configurations were added that were not chosen in the noiseless sampling results.

\subsection{QSCI and QSCI-PT results}
\begin{figure}[H]
\centering
\includegraphics[width=6.1in] {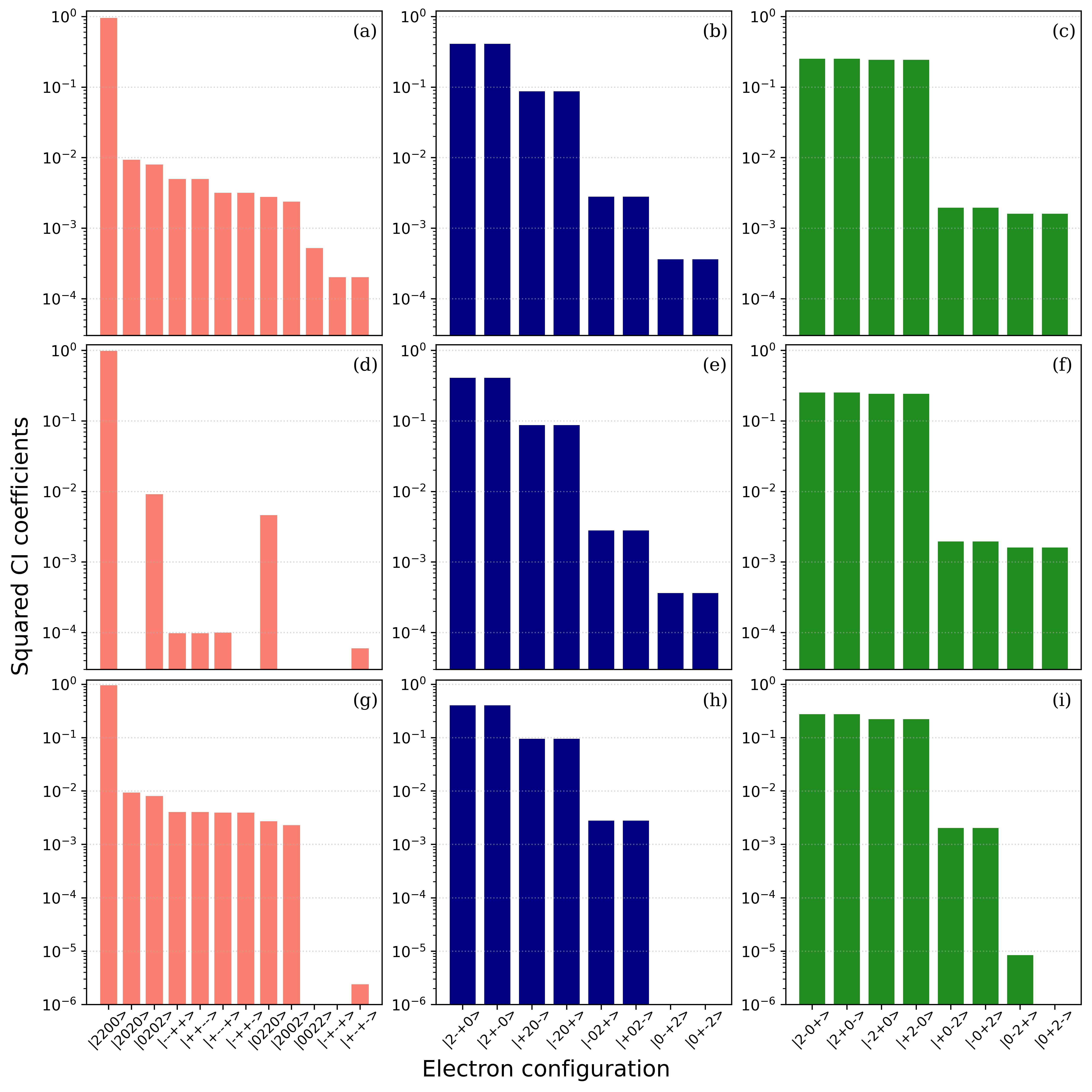}
\caption{Squared CI coefficients associated with CASCI and QSCI calculations for a naphthalene molecule.
The top row shows the results of CASCI calculations for the (a) ground state, (b) $^{1}\rm{L_{a}}$ excited state and (c) $^{1}\rm{L_{b}}$ excited state.
The middle row shows the results of noiseless QSCI simulations for the (d) ground state, (e) $^{1}\rm{L_{a}}$ excited state and (f) $^{1}\rm{L_{b}}$ excited state.
The bottom row shows the results of QSCI simulations using \texttt{ibmq\_osaka} for the (g) ground state, (h) $^{1}\rm{L_{a}}$ excited state and (i) $^{1}\rm{L_{b}}$ excited state.
The \textit{x}-axis labels indicate the electron configurations, the order of which aligns with those obtained from CASCI calculations as shown in the first row.
}
\label{fig:elec_config_casci_qsci}
\end{figure}

The electronic state energies and excitation energies for a naphthalene molecule as calculated using the CI and GMC-QDPT methods based on a configuration space with four active orbitals from the HOMO--1 to LUMO+1 are summarized in Table \ref{table:0404-results}.
It is helpful to first discuss the effects of electron configurations in detail based on the CI results.
In all cases, all the main configurations of each state shown in Figure \ref{fig:La-Lb-main-config} were included.
Thus, there is no significant difference in the energies of the electronic states.
Nevertheless, the differences in the electron configurations involved in the configuration spaces are reflected in slight variations in energy values.
A detailed analysis of these results reveals the features and advantages of the QSCI method.
Since the three states considered in this process all have different spatial symmetries and are the lowest energy states for each symmetry, the energy difference due to the configuration space used for the CI calculations can be attributed to the magnitude of the contribution of the electron configurations involved in that space.
The QSCI calculations showed that the ground state energy was slightly higher than that provided by the CISD(4e, 4o) method, whereas the $^{1}\rm{L_{a}}$ and $^{1}\rm{L_{b}}$ excited state energies were lower.
The main configuration of the ground state was the Hartree-Fock type ground configuration, which interacts primarily with doubly-excited configurations.
The number of doubly-excited configurations in the QSCI space was also found to be less than that in the CISD(4e, 4o) space, and so the estimated ground state energy was slightly higher.
Conversely, the QSCI algorithm provided energy values that were lower by 2.6 mHartree compared with the CISD(4e, 4o) values for both the $^{1}\rm{L_{a}}$ and $^{1}\rm{L_{b}}$ excited states.
These differences are ascribed to the incorporation of triply-excited configurations in the QSCI space but not in the CISD(4e, 4o) space.
The weights of the electron configurations resulting from the CI calculations shown in Figure \ref{fig:elec_config_casci_qsci} indicate that the triply-excited configurations contributed to the excited states.
As a result, the excited state energies obtained from the QSCI algorithm agreed with those provided by the CAS(4e, 4o) calculations within 0.1 mHartree.
These results suggest that using input states that include important electronic configurations allowed the QSCI algorithm to construct a configuration space of the same dimension with the CISD(4e, 4o) space but having configurations that made larger contributions to the excited states.

It is also important to examine the calculated excitation energies.
The ordering of the calculated excitation energies was found to be in agreement with the experimental results~\cite{grimme2003}. That is, in all cases, $^{1}\rm{L_{a}}$ $>$ $^{1}\rm{L_{b}}$.
However, the calculated excitation energies were significantly higher than the experimental values even when the CAS(4e, 4o) process was adopted, with deviations in excess of 1 eV.
The calculated excitation energy for the $^{1}\rm{L_{a}}$ state was also too close to that for $^{1}\rm{L_{b}}$ state, which is again inconsistent with the experimental results.
The QSCI-PT calculations provided greater accuracy and reduced the deviations from the experimental data to less than 0.5 eV for the $^{1}\rm{L_{a}}$ state and 0.8 eV for the $^{1}\rm{L_{b}}$ state.
The improvements observed in the case of the QSCI-PT calculations suggest that a lack of dynamical correlation was the primary cause of the deviations in the CI results.
Nevertheless, the QSCI-PT technique underestimated the excitation energies for both the $^{1}\rm{L_{a}}$ and $^{1}\rm{L_{b}}$ states, indicating that the effects of dynamical correlation were actually overestimated.
This outcome was not unique to the QSCI algorithm, but was also the case with the CISD(4e, 4o) and CAS(4e, 4o) calculations, suggesting that the quality of these configuration spaces as references for the GMC-QDPT method was insufficient.
The quality of the reference space could be improved by adding more electron configurations, which would require greater quantities of active orbitals and active electrons.
Because this expansion of the reference space was not possible with present-day NISQ devices, this was instead achieved using a classical computer in the present study.

The calculation results obtained with an augmented configuration space based on ten active orbitals are collected in Table \ref{table:1010-results}.
At the CI level, the excitation energy for the $^{1}\rm{L_{b}}$ state was significantly reduced compared with that for the $^{1}\rm{L_{a}}$ state.
The excitation energy for the $^{1}\rm{L_{b}}$ state approached the experimental value and the gap between this value and that for the $^{1}\rm{L_{a}}$ state increased.
The $^{1}\rm{L_{b}}$ excited state is known to exhibit more multiconfiguration character than the $^{1}\rm{L_{a}}$ excited state.~\cite{shiraiyanai2016}
Therefore, in the case of the former, the effect of the configuration space extension might be expected to be more apparent.
The excitation energies obtained from the GMC-QDPT calculations using the QSCI + X technique were in quantitatively good agreement with the experimental values as well as the results of the CAS(10e, 10o) calculations.
The deviations from the experimental data were 0.01 eV for the $^{1}\rm{L_{a}}$ state and 0.24 eV for the $^{1}\rm{L_{b}}$ state at a maximum, respectively.
Note that, as a consequence of the X excitations, all configurations that were not originally selected in the QSCI sampling due to noise effects were included.
That is, this extension of the configuration space was able to compensate for the loss of accuracy stemming from the unexpected noise-derived exclusion of electron configurations.
The dimensions of the QSCI + X configuration space compared with that of the CAS(10e, 10o) space were approximately 12\% for X = SD, 44\% for X = SDT and 81\% for X = SDTQ.
Even though the number of electron configurations associated with the QSCI + SD calculations was less than 1/8th that for the CAS(10e, 10o), the GMC-QDPT method when using X = SD allowed the excitation energy to be calculated quantitatively, suggesting that it provides a good reference for MRPT calculations. This work demonstrates that, if the electron configuration space can be truncated by using the QSCI algorithm to extract those electron configurations having larger weights, the dimension of the augmented space can be reduced.
Of course, if the QSCI method is able to handle a large number of active orbitals in the future, expansions of the configuration space based on the use of classical computers may become unnecessary.

\begin{figure}[H]
\centering
\includegraphics[width=6.1in] {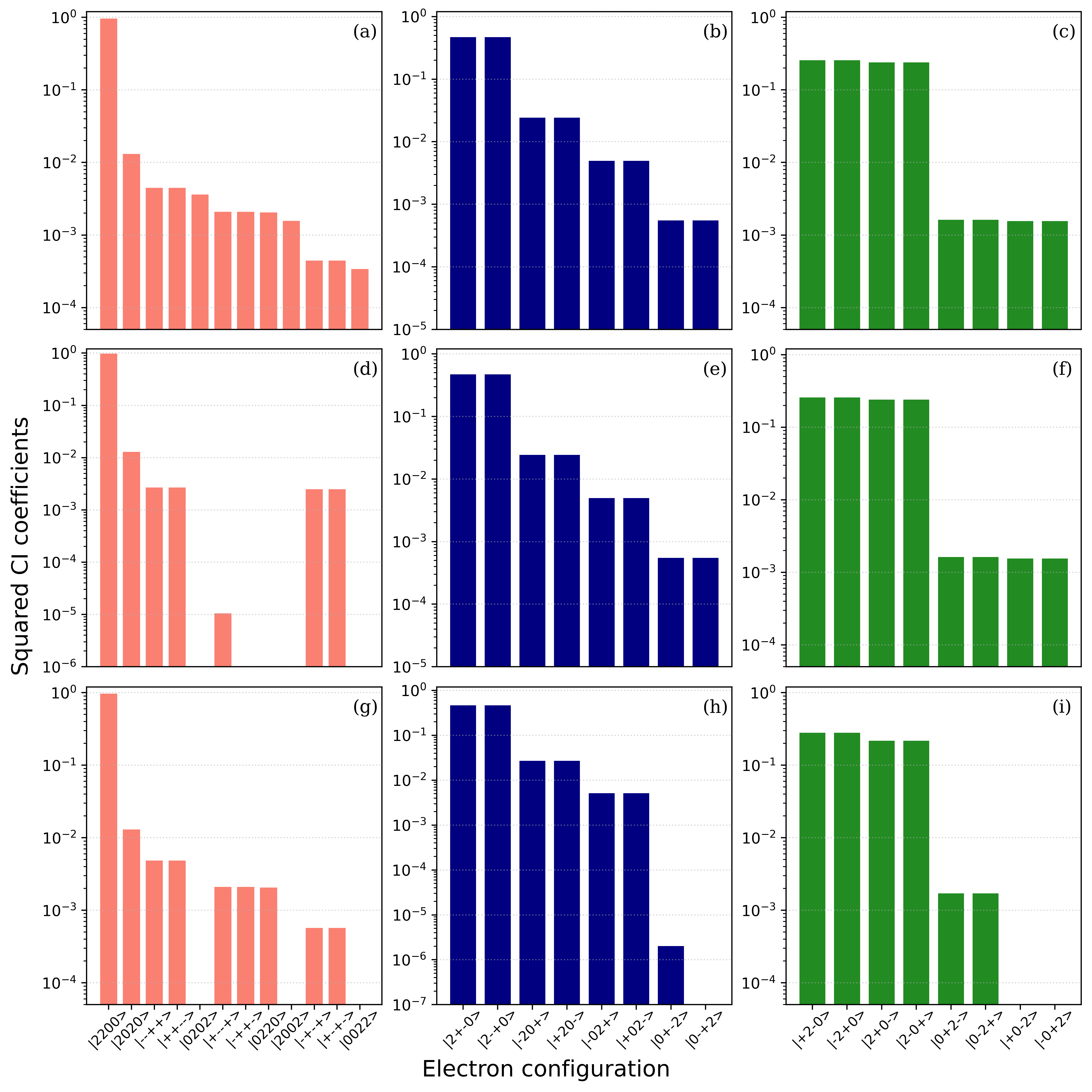}
\caption{
Squared CI coefficients associated with CASCI and QSCI calculations for a tetracene molecule.
The top row shows the results of CASCI calculations for the (a) ground state, (b)  $^{1}\rm{L_{a}}$ excited state and (c) $^{1}\rm{L_{b}}$ excited state.
The middle row shows the results of noiseless QSCI simulations for the (d) ground state, (e) $^{1}\rm{L_{a}}$ excited state and (f) $^{1}\rm{L_{b}}$ excited state.
The bottom row shows the results of QSCI simulations using \texttt{ibmq\_osaka} for the (g) ground state, (h) $^{1}\rm{L_{a}}$ excited state and (i) $^{1}\rm{L_{b}}$ excited state.
The \textit{x}-axis labels indicate the electron configurations, the order of which aligns with those obtained from CASCI calculations, as shown in the first row.
}
\label{fig:elec_config_casci_qsci_tetracene}
\end{figure}

The results of CI and GMC-QDPT calculations for a tetracene molecule are summarized in Table \ref{table:0404-1818-tetra}.
In contrast to the results obtained for naphthalene, the excitation energy for the $^{1}\rm{L_{a}}$ state obtained from the CI calculations based on four active orbitals was lower than that for the $^{1}\rm{L_{b}}$ state. This outcome was consistent with the experimental results.~\cite{grimme2003, nijegorodov1997}
Comparing the electronic state energies calculated  by the QSCI method with the CISD(4e, 4o) results, the ground state energies were higher and the excited state energies were lower.
The weights of the electron configurations in the CI results suggest that the doubly-excited configurations interacted in the case of the ground state, whereas the triply-excited configurations contributed to the excited state (Figure \ref{fig:elec_config_casci_qsci_tetracene}).
Thus, the calculation results reflect the structure of the configuration space, in that the CISD(4e, 4o) calculations included all doubly-excited configurations while the QSCI algorithm included some triply-excited configurations.
It is evident that the quantitative accuracy obtained from these calculations was insufficient. Specifically, the calculated excitation energies were more than 1 eV higher than the experimental values.
The GMC-QDPT calculations reduced this discrepancy to less than 1 eV, but underestimated the excitation energies.
These trends were also observed even for the CAS(4e, 4o) results, suggesting that an augmentation of the configuration space would be required to improve the accuracy.
In the case of the CI results based on the QSCI + SD space, the ordering of the excitation energies was unexpectedly inverted to $^{1}\rm{L_{a}}$ $>$ $^{1}\rm{L_{b}}$.
This problem was solved by applying the GMC-QDPT method, which provided excitation energies in relatively good agreement with the experimental values.
The dimension of the QSCI + SD space was 141,107 and so was extremely small given that the dimension of the CAS(18e, 18o) space was $2.36 \times 10^{9}$.
Thus, if the electron configuration space can be suitably truncated by quantum selection, the dimension of the augmented QSCI space can be reduced such that it is far less than that of the CAS space.
As noted, because the QSCI process is able to handle a greater number of active orbitals based on the use of quantum devices, such augmentation of the configuration space may become unnecessary.
Nevertheless, this augmentation could still be useful as a means of compensating for the loss of accuracy due to incorrect selection of configurations caused by quantum noise.
The results of the present QSCI/QSCI-PT calculations for naphthalene and tetracene molecules, together with the sampling results shown in Figures \ref{fig:combined_sampling_result} and \ref{fig:combined_sampling_result_tetracene},  suggest that the proposed calculation scheme is not restricted to specific molecules but rather is generally applicable to the analysis of the excited states of aromatic molecules.

One concern associated with this study was the use of the VQD approach to prepare the initial states for QSCI calculations using classical computers. Increasing the number of qubits above a certain threshold will no longer allow the simulation of these states.
It should also be noted that methods for preparing input states for QSCI and QSCI-PT calculations will continue to have associated challenges if the size of the system is scaled up.
While some proposals for QSCI input state preparation exist~\cite{Nakagawa2024, sugisaki2024, mikkelsen2025}, the development of more efficient and scalable techniques remains an open problem. Addressing these challenges is an important direction for future research.

\begin{table}[H]
\centering
    \caption{Ground and excited state energies and excitation energies for    $^{1}$L$_{\rm{a}}$ and $^{1}$L$_{\rm{b}}$ states of a naphthalene molecule calculated using the CI and GMC-QDPT processes with a configuration space based on four active orbitals from HOMO--1 to LUMO+1. Available experimental values are also shown.}
    \scriptsize
    \renewcommand{\arraystretch}{1.35}
    \begin{tabular}{llrcccccc}
    \hline
    \multicolumn{1}{c}{Method} &
    \multicolumn{1}{c}{Configuration space} & \multicolumn{1}{c}{Dimension}
    & \multicolumn{3}{c}{Electronic state energy (Hartree)}
    & & \multicolumn{2}{c}{Excitation energy (eV)} \\ \cline{4-6} \cline{8-9}
    & & & Ground state & $^{1}$L$_{\rm{a}}$ & $^{1}$L$_{\rm{b}}$
    & & $^{1}$L$_{\rm{a}}$ & $^{1}$L$_{\rm{b}}$ \\ \hline
    CI & QSCI & 27 &
    -383.4057 & -383.1763 & -383.1814 & & 6.24 & 6.10 \\
    & CISD(4e, 4o) & 27 &
    -383.4058 & -383.1737 & -383.1788 & & 6.32 & 6.18 \\
    & CAS(4e, 4o) & 36 &
    -383.4063 & -383.1766 & -383.1823 & & 6.25 & 6.10 \\ \\
    GMC-QDPT & QSCI & 27 &
    -384.6783 & -384.5194 & -384.5512 & & 4.32 & 3.46  \\
    & CISD(4e, 4o) & 27 &
    -384.6781 & -384.5225 & -384.5539 & & 4.23 & 3.38  \\
    & CAS(4e, 4o) & 36 &
    -384.6782 & -384.5201 & -384.5509 & & 4.30 & 3.46  \\
    exptl.$^{a}$ & & & & & & & 4.66 & 4.13 \\ \hline
    \multicolumn{9}{l}{$^{a}$ Reference \cite{grimme2003}}
    \end{tabular}
    \label{table:0404-results}
\end{table}

\begin{table}[H]
\centering
    \caption{Ground and excited state energies and excitation energies for $^{1}$L$_{\rm{a}}$ and $^{1}$L$_{\rm{b}}$ states calculated using the CI and GMC-QDPT processes. Results of calculations based on the CISD(4e, 4o) and CAS(4e, 4o) spaces are also provided for comparison along with available experimental data.}
    \scriptsize
    \renewcommand{\arraystretch}{1.35}
    \begin{tabular}{llrcccccc}
    \hline
    \multicolumn{1}{c}{Method} &
    \multicolumn{1}{c}{Configuration space} & \multicolumn{1}{c}{Dimension}
    & \multicolumn{3}{c}{Electronic state energy (Hartree)}
    & & \multicolumn{2}{c}{Excitation energy (eV)} \\ \cline{4-6} \cline{8-9}
    & & & Ground state & $^{1}$L$_{\rm{a}}$ & $^{1}$L$_{\rm{b}}$
    & & $^{1}$L$_{\rm{a}}$ & $^{1}$L$_{\rm{b}}$ \\ \hline
    CI & QSCI+SD  & 7641  &
    -383.4732 & -383.2507 & -383.3060 & & 6.05 & 4.55 \\
    & QSCI+SDT & 27855 &
    -383.4757 & -383.2537 & -383.3143 & & 6.04 & 4.39 \\
    & QSCI+SDTQ & 51480 &
    -383.4770 & -383.2543 & -383.3160 & & 6.06 & 4.38 \\
    & CAS(10e, 10o) & 63504 &
    -383.4771 & -383.2543 & -383.3161 & & 6.06 & 4.38 \\ \\
    GMC-QDPT & QSCI+SD  & 7641  &
    -384.6813 & -384.5103 & -384.5384 & & 4.65 & 3.89  \\
    & QSCI+SDT & 27855 &
    -384.6815 & -384.5104 & -384.5379 & & 4.66 & 3.91  \\
    & QSCI+SDTQ & 51480 &
    -384.6818 & -384.5105 & -384.5380 & & 4.66 & 3.91  \\
    & CAS(10e, 10o) & 63504 &
    -384.6818 & -384.5106 & -384.5380 & & 4.66 & 3.91  \\
    exptl.$^{a}$ & & & & & & & 4.66 & 4.13 \\ \hline
    \multicolumn{9}{l}{$^{a}$ Reference \cite{grimme2003}}
    \end{tabular}
    \label{table:1010-results}
\end{table}

\begin{table}[H]
\centering
    \caption{Ground and excited state energies and excitation energies for    $^{1}$L$_{\rm{a}}$ and $^{1}$L$_{\rm{b}}$ states of a tetracene molecule calculated using the CI and GMC-QDPT processes. Available experimental values are also shown.}
    \scriptsize
    \renewcommand{\arraystretch}{1.35}
    \begin{tabular}{llrcccccc}
    \hline
    \multicolumn{1}{c}{Method} &
    \multicolumn{1}{c}{Configuration space} & \multicolumn{1}{c}{Dimension}
    & \multicolumn{3}{c}{Electronic state energy (Hartree)}
    & & \multicolumn{2}{c}{Excitation energy (eV)} \\ \cline{4-6} \cline{8-9}
    & & & Ground state & $^{1}$L$_{\rm{a}}$ & $^{1}$L$_{\rm{b}}$
    & & $^{1}$L$_{\rm{a}}$ & $^{1}$L$_{\rm{b}}$ \\ \hline
    CI & QSCI & 27 &
    -688.6990 & -688.5463 & -688.5152 & &  4.16 & 5.00\\
       & CISD(4e, 4o) & 27 &
    -688.7014 & -688.5418 & -688.5139 & &  4.34 & 5.10\\
       & CAS(4e, 4o) & 36 &
    -688.7017 & -688.5466 & -688.5164 & &  4.22 & 5.04\\
       & QSCI + SD & 141107 &
    -688.8383 & -688.6879 & -688.6949 & &  4.09 & 3.90\\
    GMC-QDPT & QSCI & 27 &
    -691.0101 & -690.9259 & -690.9206 & & 2.29 & 2.44\\
           & CISD(4e, 4o) & 27 &
    -691.0104 & -690.9287 & -690.9214 & & 2.22 & 2.42\\
           & CAS(4e, 4o)  & 36 &
    -691.0105 & -690.9262 & -690.9200 & & 2.29 & 2.46\\
           & QSCI + SD & 141107 &
    -691.0243 & -690.9227 & -690.9146 & &  2.77 & 2.99\\
    exptl. & & & & & & & 2.88$^{a}$, 2.60$^{b}$ & 3.39$^{a}$, 3.14$^{b}$ \\ \hline
    \multicolumn{9}{l}{$^{a}$ Reference \cite{grimme2003}}\\
    \multicolumn{9}{l}{$^{b}$ Reference \cite{nijegorodov1997}}
    \end{tabular}
    \label{table:0404-1818-tetra}
\end{table}

\section{Conclusion}
QSCI is a promising algorithm that takes advantage of the ability of quantum computers to work with large-dimensional electron configuration spaces while reducing the adverse effects of noise on qubits.
As a means of compensating for the shortcomings of the QSCI process when used with present-day NISQ devices, which can only handle a small number of active orbitals due to noise, we propose the QSCI-PT method. This technique combines the QSCI algorithm with the GMC-QDPT method to improve the accuracy of calculations.
The GMC-QDPT technique is able to work with electron configuration spaces having arbitrary structures and so is easily combined with the QSCI algorithm.
The present work applied this process to the naphthalene molecule and identified the conditions necessary for the selection of electron configurations making larger contributions to the electronic state of interest. This work confirmed that the accuracy can be improved over the original QSCI by performing perturbation calculations. 
This research also demonstrated that excitation energy can be calculated with quantitative accuracy by extending the reference space based on the QSCI space. Calculations for tetracene were carried out in the same manner, confirming the versatility of the proposed calculation scheme.

Because the use of present-day quantum computers (that is, NISQ devices) remains limited to relatively small systems, the role of classical calculations in the QSCI-PT method is significant.
As the performance of quantum computers improves, it will become possible to perform calculations using a larger amount of qubits, gradually increasing the proportion of quantum computations in the QSCI-PT method.
In the future, it is likely that the benefits of quantum computing can be seamlessly experienced by using the proposed QSCI-PT approach.

\section*{Acknowledgment}
We would like to thank Dr. Keita Kanno at QunaSys for his valuable comments. We acknowledge the use of IBM Quantum services for this work. The views expressed are those of the authors, and do not reflect the official policy or position of IBM or the IBM Quantum team.

\section*{Author information}
\flushleft{
\textbf{Corresponding author}
\linebreak
\linebreak
Soichi Shirai -- Toyota Central Research and Development Laboratories, Incorporated, Nagakute, Aichi 480-1192, Japan; \url{https://orcid.org/0000-0001-6932-4845} ;
\\
Email: shirai@mosk.tytlabs.co.jp
}\linebreak

Sho koh -- QunaSys Inc., Aqua Hakusan Building 9F, 1-13-7 Hakusan, Bunkyo, Tokyo 113-0001, Japan; \url{https://orcid.org/0000-0002-8092-8997} ;
\\
Email: koh@qunasys.com
\linebreak
\linebreak
\textbf{Author}
\linebreak
\linebreak
Shih-Yen Tseng -- QunaSys Inc., Aqua Hakusan Building 9F, 1-13-7 Hakusan, Bunkyo, Tokyo 113-0001, Japan; \url{https://orcid.org/0000-0003-3942-660X}
\linebreak
\linebreak
Hokuto Iwakiri -- QunaSys Inc., Aqua Hakusan Building 9F, 1-13-7 Hakusan, Bunkyo, Tokyo 113-0001, Japan; \url{https://orcid.org/0000-0002-7316-2698}
\linebreak
\linebreak
Takahiro Horiba -- Toyota Central Research and Development Laboratories, Incorporated, Nagakute, Aichi 480-1192, Japan; \url{https://orcid.org/0000-0002-8610-047X}
\linebreak
\linebreak
Hirotoshi Hirai -- Toyota Central Research and Development Laboratories, Incorporated, Nagakute, Aichi 480-1192, Japan; \url{https://orcid.org/0000-0002-9618-3387}
\linebreak
\linebreak
\textbf{Notes}
\linebreak
\linebreak
The authors declare no competing financial interest.

\bibliography{ref}

\providecommand{\latin}[1]{#1}
\makeatletter
\providecommand{\doi}
  {\begingroup\let\do\@makeother\dospecials
  \catcode`\{=1 \catcode`\}=2 \doi@aux}
\providecommand{\doi@aux}[1]{\endgroup\texttt{#1}}
\makeatother
\providecommand*\mcitethebibliography{\thebibliography}
\csname @ifundefined\endcsname{endmcitethebibliography}  {\let\endmcitethebibliography\endthebibliography}{}
\begin{mcitethebibliography}{74}
\providecommand*\natexlab[1]{#1}
\providecommand*\mciteSetBstSublistMode[1]{}
\providecommand*\mciteSetBstMaxWidthForm[2]{}
\providecommand*\mciteBstWouldAddEndPuncttrue
  {\def\EndOfBibitem{\unskip.}}
\providecommand*\mciteBstWouldAddEndPunctfalse
  {\let\EndOfBibitem\relax}
\providecommand*\mciteSetBstMidEndSepPunct[3]{}
\providecommand*\mciteSetBstSublistLabelBeginEnd[3]{}
\providecommand*\EndOfBibitem{}
\mciteSetBstSublistMode{f}
\mciteSetBstMaxWidthForm{subitem}{(\alph{mcitesubitemcount})}
\mciteSetBstSublistLabelBeginEnd
  {\mcitemaxwidthsubitemform\space}
  {\relax}
  {\relax}

\bibitem[Jones(2015)]{jones2015density}
Jones,~R.~O. Density functional theory: Its origins, rise to prominence, and future. \emph{Rev. Mod. Phys.} \textbf{2015}, \emph{87}, 897--923\relax
\mciteBstWouldAddEndPuncttrue
\mciteSetBstMidEndSepPunct{\mcitedefaultmidpunct}
{\mcitedefaultendpunct}{\mcitedefaultseppunct}\relax
\EndOfBibitem
\bibitem[Sherrill and Schaefer~III(1999)Sherrill, and Schaefer~III]{ci1999}
Sherrill,~C.~D.; Schaefer~III,~H.~F. \emph{Adv. Quantum Chem.}; Elsevier, 1999; Vol.~34; pp 143--269\relax
\mciteBstWouldAddEndPuncttrue
\mciteSetBstMidEndSepPunct{\mcitedefaultmidpunct}
{\mcitedefaultendpunct}{\mcitedefaultseppunct}\relax
\EndOfBibitem
\bibitem[Olsen \latin{et~al.}(1996)Olsen, J{\o}rgensen, Koch, Balkova, and Bartlett]{fci1996}
Olsen,~J.; J{\o}rgensen,~P.; Koch,~H.; Balkova,~A.; Bartlett,~R.~J. Full configuration--interaction and state of the art correlation calculations on water in a valence double-zeta basis with polarization functions. \emph{J. Chem. Phys.} \textbf{1996}, \emph{104}, 8007--8015\relax
\mciteBstWouldAddEndPuncttrue
\mciteSetBstMidEndSepPunct{\mcitedefaultmidpunct}
{\mcitedefaultendpunct}{\mcitedefaultseppunct}\relax
\EndOfBibitem
\bibitem[Cao \latin{et~al.}(2019)Cao, Romero, Olson, Degroote, Johnson, Kieferov{\'a}, Kivlichan, Menke, Peropadre, Sawaya, \latin{et~al.} others]{qcqc2019}
Cao,~Y.; Romero,~J.; Olson,~J.~P.; Degroote,~M.; Johnson,~P.~D.; Kieferov{\'a},~M.; Kivlichan,~I.~D.; Menke,~T.; Peropadre,~B.; Sawaya,~N.~P.; others Quantum chemistry in the age of quantum computing. \emph{Chem. Rev.} \textbf{2019}, \emph{119}, 10856--10915\relax
\mciteBstWouldAddEndPuncttrue
\mciteSetBstMidEndSepPunct{\mcitedefaultmidpunct}
{\mcitedefaultendpunct}{\mcitedefaultseppunct}\relax
\EndOfBibitem
\bibitem[McArdle \latin{et~al.}(2020)McArdle, Endo, Aspuru-Guzik, Benjamin, and Yuan]{qcqc2020-1}
McArdle,~S.; Endo,~S.; Aspuru-Guzik,~A.; Benjamin,~S.~C.; Yuan,~X. Quantum computational chemistry. \emph{Rev. Mod. Phys.} \textbf{2020}, \emph{92}, 015003\relax
\mciteBstWouldAddEndPuncttrue
\mciteSetBstMidEndSepPunct{\mcitedefaultmidpunct}
{\mcitedefaultendpunct}{\mcitedefaultseppunct}\relax
\EndOfBibitem
\bibitem[Bauer \latin{et~al.}(2020)Bauer, Bravyi, Motta, and Chan]{qcqc2020-2}
Bauer,~B.; Bravyi,~S.; Motta,~M.; Chan,~G. K.-L. Quantum algorithms for quantum chemistry and quantum materials science. \emph{Chem. Rev.} \textbf{2020}, \emph{120}, 12685--12717\relax
\mciteBstWouldAddEndPuncttrue
\mciteSetBstMidEndSepPunct{\mcitedefaultmidpunct}
{\mcitedefaultendpunct}{\mcitedefaultseppunct}\relax
\EndOfBibitem
\bibitem[Preskill(2018)]{nisq2018}
Preskill,~J. Quantum computing in the NISQ era and beyond. \emph{Quantum} \textbf{2018}, \emph{2}, 79\relax
\mciteBstWouldAddEndPuncttrue
\mciteSetBstMidEndSepPunct{\mcitedefaultmidpunct}
{\mcitedefaultendpunct}{\mcitedefaultseppunct}\relax
\EndOfBibitem
\bibitem[Wang \latin{et~al.}(2019)Wang, Higgott, and Brierley]{nisq2019}
Wang,~D.; Higgott,~O.; Brierley,~S. Accelerated variational quantum eigensolver. \emph{Phys. Rev. Lett.} \textbf{2019}, \emph{122}, 140504\relax
\mciteBstWouldAddEndPuncttrue
\mciteSetBstMidEndSepPunct{\mcitedefaultmidpunct}
{\mcitedefaultendpunct}{\mcitedefaultseppunct}\relax
\EndOfBibitem
\bibitem[Sugisaki \latin{et~al.}(2021)Sugisaki, Sakai, Toyota, Sato, Shiomi, and Takui]{nisq2021}
Sugisaki,~K.; Sakai,~C.; Toyota,~K.; Sato,~K.; Shiomi,~D.; Takui,~T. Bayesian phase difference estimation: a general quantum algorithm for the direct calculation of energy gaps. \emph{Phys. Chem. Chem. Phys.} \textbf{2021}, \emph{23}, 20152--20162\relax
\mciteBstWouldAddEndPuncttrue
\mciteSetBstMidEndSepPunct{\mcitedefaultmidpunct}
{\mcitedefaultendpunct}{\mcitedefaultseppunct}\relax
\EndOfBibitem
\bibitem[Peruzzo \latin{et~al.}(2014)Peruzzo, McClean, Shadbolt, Yung, Zhou, Love, Aspuru-Guzik, and O'Brien]{Peruzzo:2013bzg}
Peruzzo,~A.; McClean,~J.; Shadbolt,~P.; Yung,~M.-H.; Zhou,~X.-Q.; Love,~P.~J.; Aspuru-Guzik,~A.; O'Brien,~J.~L. {A variational eigenvalue solver on a photonic quantum processor}. \emph{Nature Commun.} \textbf{2014}, \emph{5}, 4213\relax
\mciteBstWouldAddEndPuncttrue
\mciteSetBstMidEndSepPunct{\mcitedefaultmidpunct}
{\mcitedefaultendpunct}{\mcitedefaultseppunct}\relax
\EndOfBibitem
\bibitem[McClean \latin{et~al.}(2016)McClean, Romero, Babbush, and Aspuru-Guzik]{vqe2016}
McClean,~J.~R.; Romero,~J.; Babbush,~R.; Aspuru-Guzik,~A. The theory of variational hybrid quantum-classical algorithms. \emph{New J. Phys.} \textbf{2016}, \emph{18}, 023023\relax
\mciteBstWouldAddEndPuncttrue
\mciteSetBstMidEndSepPunct{\mcitedefaultmidpunct}
{\mcitedefaultendpunct}{\mcitedefaultseppunct}\relax
\EndOfBibitem
\bibitem[Tilly \latin{et~al.}(2022)Tilly, Chen, Cao, Picozzi, Setia, Li, Grant, Wossnig, Rungger, Booth, and Tennyson]{Tilly:2021jem}
Tilly,~J.; Chen,~H.; Cao,~S.; Picozzi,~D.; Setia,~K.; Li,~Y.; Grant,~E.; Wossnig,~L.; Rungger,~I.; Booth,~G.~H.; Tennyson,~J. The Variational Quantum Eigensolver: A review of methods and best practices. \emph{Phys. Rep.} \textbf{2022}, \emph{986}, 1--128, The variational quantum eigensolver: a review of methods and best practices\relax
\mciteBstWouldAddEndPuncttrue
\mciteSetBstMidEndSepPunct{\mcitedefaultmidpunct}
{\mcitedefaultendpunct}{\mcitedefaultseppunct}\relax
\EndOfBibitem
\bibitem[Fedorov \latin{et~al.}(2022)Fedorov, Peng, Govind, and Alexeev]{vqe2022fedorov}
Fedorov,~D.~A.; Peng,~B.; Govind,~N.; Alexeev,~Y. VQE method: a short survey and recent developments. \emph{Mater. Theory} \textbf{2022}, \emph{6}, 2\relax
\mciteBstWouldAddEndPuncttrue
\mciteSetBstMidEndSepPunct{\mcitedefaultmidpunct}
{\mcitedefaultendpunct}{\mcitedefaultseppunct}\relax
\EndOfBibitem
\bibitem[Kanno \latin{et~al.}(2023)Kanno, Kohda, Imai, Koh, Mitarai, Mizukami, and Nakagawa]{Kanno:2023rfr}
Kanno,~K.; Kohda,~M.; Imai,~R.; Koh,~S.; Mitarai,~K.; Mizukami,~W.; Nakagawa,~Y.~O. Quantum-Selected Configuration Interaction: classical diagonalization of Hamiltonians in subspaces selected by quantum computers. 2023; \url{https://arxiv.org/abs/2302.11320}\relax
\mciteBstWouldAddEndPuncttrue
\mciteSetBstMidEndSepPunct{\mcitedefaultmidpunct}
{\mcitedefaultendpunct}{\mcitedefaultseppunct}\relax
\EndOfBibitem
\bibitem[Nakano \latin{et~al.}(2002)Nakano, Uchiyama, and Hirao]{nakano2002}
Nakano,~H.; Uchiyama,~R.; Hirao,~K. Quasi-degenerate perturbation theory with general multiconfiguration self-consistent field reference functions. \emph{J. Comput. Chem.} \textbf{2002}, \emph{23}, 1166--1175\relax
\mciteBstWouldAddEndPuncttrue
\mciteSetBstMidEndSepPunct{\mcitedefaultmidpunct}
{\mcitedefaultendpunct}{\mcitedefaultseppunct}\relax
\EndOfBibitem
\bibitem[Sun \latin{et~al.}(2022)Sun, Endo, Lin, Hayden, Vedral, and Yuan]{PhysRevLett.129.120505}
Sun,~J.; Endo,~S.; Lin,~H.; Hayden,~P.; Vedral,~V.; Yuan,~X. Perturbative Quantum Simulation. \emph{Phys. Rev. Lett.} \textbf{2022}, \emph{129}, 120505\relax
\mciteBstWouldAddEndPuncttrue
\mciteSetBstMidEndSepPunct{\mcitedefaultmidpunct}
{\mcitedefaultendpunct}{\mcitedefaultseppunct}\relax
\EndOfBibitem
\bibitem[Ryabinkin \latin{et~al.}(2021)Ryabinkin, Izmaylov, and Genin]{ryabinkin2021posteriori}
Ryabinkin,~I.~G.; Izmaylov,~A.~F.; Genin,~S.~N. A posteriori corrections to the iterative qubit coupled cluster method to minimize the use of quantum resources in large-scale calculations. \emph{Quantum Sci. Technol.} \textbf{2021}, \emph{6}, 024012\relax
\mciteBstWouldAddEndPuncttrue
\mciteSetBstMidEndSepPunct{\mcitedefaultmidpunct}
{\mcitedefaultendpunct}{\mcitedefaultseppunct}\relax
\EndOfBibitem
\bibitem[Tammaro \latin{et~al.}(2023)Tammaro, Galli, Rice, and Motta]{tammaro2023n}
Tammaro,~A.; Galli,~D.~E.; Rice,~J.~E.; Motta,~M. N-electron valence perturbation theory with reference wave functions from quantum computing: application to the relative stability of hydroxide anion and hydroxyl radical. \emph{J. Phys. Chem. A} \textbf{2023}, \emph{127}, 817--827\relax
\mciteBstWouldAddEndPuncttrue
\mciteSetBstMidEndSepPunct{\mcitedefaultmidpunct}
{\mcitedefaultendpunct}{\mcitedefaultseppunct}\relax
\EndOfBibitem
\bibitem[Li \latin{et~al.}(2023)Li, Jones, and Kais]{li2023toward}
Li,~J.; Jones,~B.~A.; Kais,~S. Toward perturbation theory methods on a quantum computer. \emph{Sci. Adv.} \textbf{2023}, \emph{9}, eadg4576\relax
\mciteBstWouldAddEndPuncttrue
\mciteSetBstMidEndSepPunct{\mcitedefaultmidpunct}
{\mcitedefaultendpunct}{\mcitedefaultseppunct}\relax
\EndOfBibitem
\bibitem[Liu \latin{et~al.}(2024)Liu, Li, and Yang]{liu2024perturbative}
Liu,~J.; Li,~Z.; Yang,~J. Perturbative variational quantum algorithms for material simulations. \emph{Electron. Struct.} \textbf{2024}, \emph{6}, 015007\relax
\mciteBstWouldAddEndPuncttrue
\mciteSetBstMidEndSepPunct{\mcitedefaultmidpunct}
{\mcitedefaultendpunct}{\mcitedefaultseppunct}\relax
\EndOfBibitem
\bibitem[Di~Paola \latin{et~al.}(2024)Di~Paola, Plekhanov, Krompiec, Kumar, Marsili, Du, Weber, Krauser, Shishenina, and Mu{\~n}oz~Ramo]{di2024platinum}
Di~Paola,~C.; Plekhanov,~E.; Krompiec,~M.; Kumar,~C.; Marsili,~E.; Du,~F.; Weber,~D.; Krauser,~J.~S.; Shishenina,~E.; Mu{\~n}oz~Ramo,~D. Platinum-based catalysts for oxygen reduction reaction simulated with a quantum computer. \emph{npj Comput. Mater.} \textbf{2024}, \emph{10}, 285\relax
\mciteBstWouldAddEndPuncttrue
\mciteSetBstMidEndSepPunct{\mcitedefaultmidpunct}
{\mcitedefaultendpunct}{\mcitedefaultseppunct}\relax
\EndOfBibitem
\bibitem[Tang and VanSlyke(1987)Tang, and VanSlyke]{el1987}
Tang,~C.~W.; VanSlyke,~S.~A. Organic electroluminescent diodes. \emph{Appl. Phys. Lett} \textbf{1987}, \emph{51}, 913--915\relax
\mciteBstWouldAddEndPuncttrue
\mciteSetBstMidEndSepPunct{\mcitedefaultmidpunct}
{\mcitedefaultendpunct}{\mcitedefaultseppunct}\relax
\EndOfBibitem
\bibitem[Greenham \latin{et~al.}(1993)Greenham, Moratti, Bradley, Friend, and Holmes]{el1993}
Greenham,~N.; Moratti,~S.; Bradley,~D.; Friend,~R.; Holmes,~A. Efficient light-emitting diodes based on polymers with high electron affinities. \emph{Nature} \textbf{1993}, \emph{365}, 628--630\relax
\mciteBstWouldAddEndPuncttrue
\mciteSetBstMidEndSepPunct{\mcitedefaultmidpunct}
{\mcitedefaultendpunct}{\mcitedefaultseppunct}\relax
\EndOfBibitem
\bibitem[Friend \latin{et~al.}(1999)Friend, Gymer, Holmes, Burroughes, Marks, Taliani, Bradley, Santos, Bredas, L{\"o}gdlund, \latin{et~al.} others]{el1999}
Friend,~R.~H.; Gymer,~R.; Holmes,~A.; Burroughes,~J.; Marks,~R.; Taliani,~C.; Bradley,~D.; Santos,~D.~D.; Bredas,~J.-L.; L{\"o}gdlund,~M.; others Electroluminescence in conjugated polymers. \emph{Nature} \textbf{1999}, \emph{397}, 121--128\relax
\mciteBstWouldAddEndPuncttrue
\mciteSetBstMidEndSepPunct{\mcitedefaultmidpunct}
{\mcitedefaultendpunct}{\mcitedefaultseppunct}\relax
\EndOfBibitem
\bibitem[G{\"u}nes \latin{et~al.}(2007)G{\"u}nes, Neugebauer, and Sariciftci]{solar-cell-2007}
G{\"u}nes,~S.; Neugebauer,~H.; Sariciftci,~N.~S. Conjugated polymer-based organic solar cells. \emph{Chem. Rev.} \textbf{2007}, \emph{107}, 1324--1338\relax
\mciteBstWouldAddEndPuncttrue
\mciteSetBstMidEndSepPunct{\mcitedefaultmidpunct}
{\mcitedefaultendpunct}{\mcitedefaultseppunct}\relax
\EndOfBibitem
\bibitem[Hains \latin{et~al.}(2010)Hains, Liang, Woodhouse, and Gregg]{solar-cell-2010-1}
Hains,~A.~W.; Liang,~Z.; Woodhouse,~M.~A.; Gregg,~B.~A. Molecular semiconductors in organic photovoltaic cells. \emph{Chem. Rev.} \textbf{2010}, \emph{110}, 6689--6735\relax
\mciteBstWouldAddEndPuncttrue
\mciteSetBstMidEndSepPunct{\mcitedefaultmidpunct}
{\mcitedefaultendpunct}{\mcitedefaultseppunct}\relax
\EndOfBibitem
\bibitem[Clarke and Durrant(2010)Clarke, and Durrant]{solar-cell-2010-2}
Clarke,~T.~M.; Durrant,~J.~R. Charge photogeneration in organic solar cells. \emph{Chem. Rev.} \textbf{2010}, \emph{110}, 6736--6767\relax
\mciteBstWouldAddEndPuncttrue
\mciteSetBstMidEndSepPunct{\mcitedefaultmidpunct}
{\mcitedefaultendpunct}{\mcitedefaultseppunct}\relax
\EndOfBibitem
\bibitem[Coropceanu \latin{et~al.}(2007)Coropceanu, Cornil, da~Silva~Filho, Olivier, Silbey, and Br{\'e}das]{hole-transport-2007}
Coropceanu,~V.; Cornil,~J.; da~Silva~Filho,~D.~A.; Olivier,~Y.; Silbey,~R.; Br{\'e}das,~J.-L. Charge transport in organic semiconductors. \emph{Chem. Rev.} \textbf{2007}, \emph{107}, 926--952\relax
\mciteBstWouldAddEndPuncttrue
\mciteSetBstMidEndSepPunct{\mcitedefaultmidpunct}
{\mcitedefaultendpunct}{\mcitedefaultseppunct}\relax
\EndOfBibitem
\bibitem[Anthony \latin{et~al.}(2010)Anthony, Facchetti, Heeney, Marder, and Zhan]{electron-trasnport-2010}
Anthony,~J.~E.; Facchetti,~A.; Heeney,~M.; Marder,~S.~R.; Zhan,~X. n-Type organic semiconductors in organic electronics. \emph{Adv. Mater.} \textbf{2010}, \emph{22}, 3876--3892\relax
\mciteBstWouldAddEndPuncttrue
\mciteSetBstMidEndSepPunct{\mcitedefaultmidpunct}
{\mcitedefaultendpunct}{\mcitedefaultseppunct}\relax
\EndOfBibitem
\bibitem[{Fock}(1932)]{fock:1932}
{Fock},~V. {Konfigurationsraum und zweite Quantelung}. \emph{Zeitschrift fur Physik} \textbf{1932}, \emph{75}, 622--647\relax
\mciteBstWouldAddEndPuncttrue
\mciteSetBstMidEndSepPunct{\mcitedefaultmidpunct}
{\mcitedefaultendpunct}{\mcitedefaultseppunct}\relax
\EndOfBibitem
\bibitem[Jordan and Wigner(1928)Jordan, and Wigner]{Jordan1928berDP}
Jordan,~P.; Wigner,~E. {\"U}ber das Paulische {\"A}quivalenzverbot. \emph{Z. Phys.} \textbf{1928}, \emph{47}, 631--651\relax
\mciteBstWouldAddEndPuncttrue
\mciteSetBstMidEndSepPunct{\mcitedefaultmidpunct}
{\mcitedefaultendpunct}{\mcitedefaultseppunct}\relax
\EndOfBibitem
\bibitem[Nielsen \latin{et~al.}(2005)Nielsen, \latin{et~al.} others]{Nielsen2005TheFC}
Nielsen,~M.~A.; others The Fermionic canonical commutation relations and the Jordan-Wigner transform. \emph{School of Physical Sciences The University of Queensland} \textbf{2005}, \emph{59}, 75\relax
\mciteBstWouldAddEndPuncttrue
\mciteSetBstMidEndSepPunct{\mcitedefaultmidpunct}
{\mcitedefaultendpunct}{\mcitedefaultseppunct}\relax
\EndOfBibitem
\bibitem[Levine \latin{et~al.}(2021)Levine, Durden, Esch, Liang, and Shu]{casci2021}
Levine,~B.~G.; Durden,~A.~S.; Esch,~M.~P.; Liang,~F.; Shu,~Y. CAS without SCF—Why to use CASCI and where to get the orbitals. \emph{J. Chem. Phys.} \textbf{2021}, \emph{154}, 090902\relax
\mciteBstWouldAddEndPuncttrue
\mciteSetBstMidEndSepPunct{\mcitedefaultmidpunct}
{\mcitedefaultendpunct}{\mcitedefaultseppunct}\relax
\EndOfBibitem
\bibitem[Roos \latin{et~al.}(1980)Roos, Taylor, and Sigbahn]{casscf1980}
Roos,~B.~O.; Taylor,~P.~R.; Sigbahn,~P.~E. A complete active space SCF method (CASSCF) using a density matrix formulated super-CI approach. \emph{Chem. Phys.} \textbf{1980}, \emph{48}, 157--173\relax
\mciteBstWouldAddEndPuncttrue
\mciteSetBstMidEndSepPunct{\mcitedefaultmidpunct}
{\mcitedefaultendpunct}{\mcitedefaultseppunct}\relax
\EndOfBibitem
\bibitem[Platt(1949)]{platt1949}
Platt,~J.~R. Classification of spectra of cata-condensed hydrocarbons. \emph{J. Chem. Phys.} \textbf{1949}, \emph{17}, 484--495\relax
\mciteBstWouldAddEndPuncttrue
\mciteSetBstMidEndSepPunct{\mcitedefaultmidpunct}
{\mcitedefaultendpunct}{\mcitedefaultseppunct}\relax
\EndOfBibitem
\bibitem[Rubio \latin{et~al.}(1994)Rubio, Merch{\'a}n, Ort{\'\i}, and Roos]{roos1994}
Rubio,~M.; Merch{\'a}n,~M.; Ort{\'\i},~E.; Roos,~B.~O. A theoretical study of the electronic spectrum of naphthalene. \emph{Chem. Phys.} \textbf{1994}, \emph{179}, 395--409\relax
\mciteBstWouldAddEndPuncttrue
\mciteSetBstMidEndSepPunct{\mcitedefaultmidpunct}
{\mcitedefaultendpunct}{\mcitedefaultseppunct}\relax
\EndOfBibitem
\bibitem[Hashimoto \latin{et~al.}(1996)Hashimoto, Nakano, and Hirao]{hirao1996}
Hashimoto,~T.; Nakano,~H.; Hirao,~K. Theoretical study of the valence $\pi$→ $\pi$* excited states of polyacenes: Benzene and naphthalene. \emph{J. Chem. Phys.} \textbf{1996}, \emph{104}, 6244--6258\relax
\mciteBstWouldAddEndPuncttrue
\mciteSetBstMidEndSepPunct{\mcitedefaultmidpunct}
{\mcitedefaultendpunct}{\mcitedefaultseppunct}\relax
\EndOfBibitem
\bibitem[Grimme and Parac(2003)Grimme, and Parac]{grimme2003}
Grimme,~S.; Parac,~M. Substantial errors from time-dependent density functional theory for the calculation of excited states of large $\pi$ systems. \emph{ChemPhysChem} \textbf{2003}, \emph{4}, 292--295\relax
\mciteBstWouldAddEndPuncttrue
\mciteSetBstMidEndSepPunct{\mcitedefaultmidpunct}
{\mcitedefaultendpunct}{\mcitedefaultseppunct}\relax
\EndOfBibitem
\bibitem[Kawashima \latin{et~al.}(1999)Kawashima, Hashimoto, Nakano, and Hirao]{kawashima1999}
Kawashima,~Y.; Hashimoto,~T.; Nakano,~H.; Hirao,~K. Theoretical study of the valence $\pi$→ $\pi$* excited states of polyacenes: anthracene and naphthacene. \emph{Theor. Chem. Acc.} \textbf{1999}, \emph{102}, 49--64\relax
\mciteBstWouldAddEndPuncttrue
\mciteSetBstMidEndSepPunct{\mcitedefaultmidpunct}
{\mcitedefaultendpunct}{\mcitedefaultseppunct}\relax
\EndOfBibitem
\bibitem[Vosko \latin{et~al.}(1980)Vosko, Wilk, and Nusair]{b3lyp1980}
Vosko,~S.~H.; Wilk,~L.; Nusair,~M. Accurate spin-dependent electron liquid correlation energies for local spin density calculations: a critical analysis. \emph{Can. J. Phys.} \textbf{1980}, \emph{58}, 1200--1211\relax
\mciteBstWouldAddEndPuncttrue
\mciteSetBstMidEndSepPunct{\mcitedefaultmidpunct}
{\mcitedefaultendpunct}{\mcitedefaultseppunct}\relax
\EndOfBibitem
\bibitem[Lee \latin{et~al.}(1988)Lee, Yang, and Parr]{b3lyp1988}
Lee,~C.; Yang,~W.; Parr,~R.~G. Development of the Colle-Salvetti correlation-energy formula into a functional of the electron density. \emph{Phys. Rev. B} \textbf{1988}, \emph{37}, 785--789\relax
\mciteBstWouldAddEndPuncttrue
\mciteSetBstMidEndSepPunct{\mcitedefaultmidpunct}
{\mcitedefaultendpunct}{\mcitedefaultseppunct}\relax
\EndOfBibitem
\bibitem[Becke(1993)]{b3lyp1993}
Becke,~A.~D. Density‐functional thermochemistry. III. The role of exact exchange. \emph{J. Chem. Phys.} \textbf{1993}, \emph{98}, 5648--5652\relax
\mciteBstWouldAddEndPuncttrue
\mciteSetBstMidEndSepPunct{\mcitedefaultmidpunct}
{\mcitedefaultendpunct}{\mcitedefaultseppunct}\relax
\EndOfBibitem
\bibitem[Stephens \latin{et~al.}(1994)Stephens, Devlin, Chabalowski, and Frisch]{b3lyp1994}
Stephens,~P.~J.; Devlin,~F.~J.; Chabalowski,~C.~F.; Frisch,~M.~J. Ab initio calculation of vibrational absorption and circular dichroism spectra using density functional force fields. \emph{J. Phys. Chem.} \textbf{1994}, \emph{98}, 11623--11627\relax
\mciteBstWouldAddEndPuncttrue
\mciteSetBstMidEndSepPunct{\mcitedefaultmidpunct}
{\mcitedefaultendpunct}{\mcitedefaultseppunct}\relax
\EndOfBibitem
\bibitem[Ditchfield \latin{et~al.}(1971)Ditchfield, Hehre, and Pople]{631g1971}
Ditchfield,~R.; Hehre,~W.~J.; Pople,~J.~A. Self-consistent molecular-orbital methods. IX. An extended Gaussian-type basis for molecular-orbital studies of organic molecules. \emph{J. Chem. Phys.} \textbf{1971}, \emph{54}, 724--728\relax
\mciteBstWouldAddEndPuncttrue
\mciteSetBstMidEndSepPunct{\mcitedefaultmidpunct}
{\mcitedefaultendpunct}{\mcitedefaultseppunct}\relax
\EndOfBibitem
\bibitem[Hehre \latin{et~al.}(1972)Hehre, Ditchfield, and Pople]{631g1972}
Hehre,~W.~J.; Ditchfield,~R.; Pople,~J.~A. Self—consistent molecular orbital methods. XII. Further extensions of Gaussian—type basis sets for use in molecular orbital studies of organic molecules. \emph{J. Chem. Phys.} \textbf{1972}, \emph{56}, 2257--2261\relax
\mciteBstWouldAddEndPuncttrue
\mciteSetBstMidEndSepPunct{\mcitedefaultmidpunct}
{\mcitedefaultendpunct}{\mcitedefaultseppunct}\relax
\EndOfBibitem
\bibitem[Hariharan and Pople(1973)Hariharan, and Pople]{631gd}
Hariharan,~P.~C.; Pople,~J.~A. The influence of polarization functions on molecular orbital hydrogenation energies. \emph{Theor. Chem. Acc.} \textbf{1973}, \emph{28}, 213--222\relax
\mciteBstWouldAddEndPuncttrue
\mciteSetBstMidEndSepPunct{\mcitedefaultmidpunct}
{\mcitedefaultendpunct}{\mcitedefaultseppunct}\relax
\EndOfBibitem
\bibitem[Shirai \latin{et~al.}(2016)Shirai, Kurashige, and Yanai]{shiraiyanai2016}
Shirai,~S.; Kurashige,~Y.; Yanai,~T. Computational evidence of inversion of $^{1}\rm{L_{a}}$ and $^{1}\rm{L_{b}}$-derived excited states in naphthalene excimer formation from ab initio multireference theory with large active space: DMRG-CASPT2 Study. \emph{J. Chem. Theory Comput.} \textbf{2016}, \emph{12}, 2366--2372\relax
\mciteBstWouldAddEndPuncttrue
\mciteSetBstMidEndSepPunct{\mcitedefaultmidpunct}
{\mcitedefaultendpunct}{\mcitedefaultseppunct}\relax
\EndOfBibitem
\bibitem[Baba \latin{et~al.}(2011)Baba, Kowaka, Nagashima, Ishimoto, Goto, and Nakayama]{nph-geom}
Baba,~M.; Kowaka,~Y.; Nagashima,~U.; Ishimoto,~T.; Goto,~H.; Nakayama,~N. Geometrical structure of benzene and naphthalene: Ultrahigh-resolution laser spectroscopy and ab initio calculation. \emph{J. Chem. Phys.} \textbf{2011}, \emph{135}, 054305\relax
\mciteBstWouldAddEndPuncttrue
\mciteSetBstMidEndSepPunct{\mcitedefaultmidpunct}
{\mcitedefaultendpunct}{\mcitedefaultseppunct}\relax
\EndOfBibitem
\bibitem[Campbell \latin{et~al.}(1962)Campbell, Robertson, and Trotter]{tetra-bonds}
Campbell,~R.; Robertson,~J.~M.; Trotter,~J. The crystal structure of hexacene, and a revision of the crystallographic data for tetracene. \emph{Acta Crystallogr.} \textbf{1962}, \emph{15}, 289--290\relax
\mciteBstWouldAddEndPuncttrue
\mciteSetBstMidEndSepPunct{\mcitedefaultmidpunct}
{\mcitedefaultendpunct}{\mcitedefaultseppunct}\relax
\EndOfBibitem
\bibitem[Robertson \latin{et~al.}(1961)Robertson, Sinclair, and Trotter]{tetra-angles}
Robertson,~J.~M.; Sinclair,~V.; Trotter,~J. The crystal and molecular structure of tetracene. \emph{Acta Crystallogr.} \textbf{1961}, \emph{14}, 697--704\relax
\mciteBstWouldAddEndPuncttrue
\mciteSetBstMidEndSepPunct{\mcitedefaultmidpunct}
{\mcitedefaultendpunct}{\mcitedefaultseppunct}\relax
\EndOfBibitem
\bibitem[Frisch \latin{et~al.}(2016)Frisch, Trucks, Schlegel, Scuseria, Robb, Cheeseman, Scalmani, Barone, Petersson, Nakatsuji, \latin{et~al.} others]{g16}
Frisch,~M.~J.; Trucks,~G.~W.; Schlegel,~H.~B.; Scuseria,~G.~E.; Robb,~M.~A.; Cheeseman,~J.~R.; Scalmani,~G.; Barone,~V.; Petersson,~G.~A.; Nakatsuji,~H.; others Gaussian 16 {R}evision {C}.01. 2016; Gaussian Inc. Wallingford CT\relax
\mciteBstWouldAddEndPuncttrue
\mciteSetBstMidEndSepPunct{\mcitedefaultmidpunct}
{\mcitedefaultendpunct}{\mcitedefaultseppunct}\relax
\EndOfBibitem
\bibitem[Dunning~Jr(1989)]{dunning1989}
Dunning~Jr,~T.~H. Gaussian basis sets for use in correlated molecular calculations. I. The atoms boron through neon and hydrogen. \emph{J. Chem. Phys.} \textbf{1989}, \emph{90}, 1007--1023\relax
\mciteBstWouldAddEndPuncttrue
\mciteSetBstMidEndSepPunct{\mcitedefaultmidpunct}
{\mcitedefaultendpunct}{\mcitedefaultseppunct}\relax
\EndOfBibitem
\bibitem[QunaSys(2024)]{quri2024}
QunaSys QURI Parts. 2024; \url{https://quri-parts.qunasys.com/}\relax
\mciteBstWouldAddEndPuncttrue
\mciteSetBstMidEndSepPunct{\mcitedefaultmidpunct}
{\mcitedefaultendpunct}{\mcitedefaultseppunct}\relax
\EndOfBibitem
\bibitem[Javadi-Abhari \latin{et~al.}(2024)Javadi-Abhari, Treinish, Krsulich, Wood, Lishman, Gacon, Martiel, Nation, Bishop, Cross, Johnson, and Gambetta]{qiskit2024}
Javadi-Abhari,~A.; Treinish,~M.; Krsulich,~K.; Wood,~C.~J.; Lishman,~J.; Gacon,~J.; Martiel,~S.; Nation,~P.~D.; Bishop,~L.~S.; Cross,~A.~W.; Johnson,~B.~R.; Gambetta,~J.~M. Quantum computing with {Q}iskit. 2024; \url{https://doi.org/10.48550/arXiv.2405.08810}\relax
\mciteBstWouldAddEndPuncttrue
\mciteSetBstMidEndSepPunct{\mcitedefaultmidpunct}
{\mcitedefaultendpunct}{\mcitedefaultseppunct}\relax
\EndOfBibitem
\bibitem[Schmidt \latin{et~al.}(1993)Schmidt, Baldridge, Boatz, Elbert, Gordon, Jensen, Koseki, Matsunaga, Nguyen, Su, \latin{et~al.} others]{gamess1993}
Schmidt,~M.~W.; Baldridge,~K.~K.; Boatz,~J.~A.; Elbert,~S.~T.; Gordon,~M.~S.; Jensen,~J.~H.; Koseki,~S.; Matsunaga,~N.; Nguyen,~K.~A.; Su,~S.; others General atomic and molecular electronic structure system. \emph{J. Comput. Chem.} \textbf{1993}, \emph{14}, 1347--1363\relax
\mciteBstWouldAddEndPuncttrue
\mciteSetBstMidEndSepPunct{\mcitedefaultmidpunct}
{\mcitedefaultendpunct}{\mcitedefaultseppunct}\relax
\EndOfBibitem
\bibitem[Gordon and Schmidt(2005)Gordon, and Schmidt]{gamess2005}
Gordon,~M.~S.; Schmidt,~M.~W. \emph{Theory and applications of computational chemistry}; Elsevier, 2005; pp 1167--1189\relax
\mciteBstWouldAddEndPuncttrue
\mciteSetBstMidEndSepPunct{\mcitedefaultmidpunct}
{\mcitedefaultendpunct}{\mcitedefaultseppunct}\relax
\EndOfBibitem
\bibitem[Barca \latin{et~al.}(2020)Barca, Bertoni, Carrington, Datta, De~Silva, Deustua, Fedorov, Gour, Gunina, Guidez, Harville, Irle, Ivanic, Kowalski, Leang, Li, Li, Lutz, Magoulas, Mato, Mironov, Nakata, Pham, Piecuch, Poole, Pruitt, Rendell, Roskop, Ruedenberg, Sattasathuchana, Schmidt, Shen, Slipchenko, Sosonkina, Sundriyal, Tiwari, Galvez~Vallejo, Westheimer, Wloch, Xu, Zahariev, and Gordon]{gamess2020}
Barca,~G. M.~J. \latin{et~al.}  Recent developments in the general atomic and molecular electronic structure system. \emph{J. Chem. Phys.} \textbf{2020}, \emph{152}, 154102\relax
\mciteBstWouldAddEndPuncttrue
\mciteSetBstMidEndSepPunct{\mcitedefaultmidpunct}
{\mcitedefaultendpunct}{\mcitedefaultseppunct}\relax
\EndOfBibitem
\bibitem[Higgott \latin{et~al.}(2019)Higgott, Wang, and Brierley]{Higgott:2018doo}
Higgott,~O.; Wang,~D.; Brierley,~S. {Variational quantum computation of excited states}. \emph{Quantum} \textbf{2019}, \emph{3}, 156\relax
\mciteBstWouldAddEndPuncttrue
\mciteSetBstMidEndSepPunct{\mcitedefaultmidpunct}
{\mcitedefaultendpunct}{\mcitedefaultseppunct}\relax
\EndOfBibitem
\bibitem[Kuroiwa and Nakagawa(2021)Kuroiwa, and Nakagawa]{Kuroiwa2021}
Kuroiwa,~K.; Nakagawa,~Y.~O. Penalty methods for a variational quantum eigensolver. \emph{Phys. Rev. Res.} \textbf{2021}, \emph{3}, 013197\relax
\mciteBstWouldAddEndPuncttrue
\mciteSetBstMidEndSepPunct{\mcitedefaultmidpunct}
{\mcitedefaultendpunct}{\mcitedefaultseppunct}\relax
\EndOfBibitem
\bibitem[Shirai \latin{et~al.}(2022)Shirai, Horiba, and Hirai]{Shirai2022}
Shirai,~S.; Horiba,~T.; Hirai,~H. Calculation of core-excited and core-ionized states using variational quantum deflation method and applications to photocatalyst modeling. \emph{ACS Omega} \textbf{2022}, \emph{7}, 10840--10853\relax
\mciteBstWouldAddEndPuncttrue
\mciteSetBstMidEndSepPunct{\mcitedefaultmidpunct}
{\mcitedefaultendpunct}{\mcitedefaultseppunct}\relax
\EndOfBibitem
\bibitem[Gard \latin{et~al.}(2020)Gard, Zhu, Barron, Mayhall, Economou, and Barnes]{gard2020efficient}
Gard,~B.~T.; Zhu,~L.; Barron,~G.~S.; Mayhall,~N.~J.; Economou,~S.~E.; Barnes,~E. Efficient symmetry-preserving state preparation circuits for the variational quantum eigensolver algorithm. \emph{npj Quantum Inf.} \textbf{2020}, \emph{6}, 10\relax
\mciteBstWouldAddEndPuncttrue
\mciteSetBstMidEndSepPunct{\mcitedefaultmidpunct}
{\mcitedefaultendpunct}{\mcitedefaultseppunct}\relax
\EndOfBibitem
\bibitem[Ibe \latin{et~al.}(2022)Ibe, Nakagawa, Earnest, Yamamoto, Mitarai, Gao, and Kobayashi]{ibe2022}
Ibe,~Y.; Nakagawa,~Y.~O.; Earnest,~N.; Yamamoto,~T.; Mitarai,~K.; Gao,~Q.; Kobayashi,~T. Calculating transition amplitudes by variational quantum deflation. \emph{Phys. Rev. Res.} \textbf{2022}, \emph{4}, 013173\relax
\mciteBstWouldAddEndPuncttrue
\mciteSetBstMidEndSepPunct{\mcitedefaultmidpunct}
{\mcitedefaultendpunct}{\mcitedefaultseppunct}\relax
\EndOfBibitem
\bibitem[Virtanen \latin{et~al.}(2020)Virtanen, Gommers, Oliphant, Haberland, Reddy, Cournapeau, Burovski, Peterson, Weckesser, Bright, {van der Walt}, Brett, Wilson, Millman, Mayorov, Nelson, Jones, Kern, Larson, Carey, Polat, Feng, Moore, {VanderPlas}, Laxalde, Perktold, Cimrman, Henriksen, Quintero, Harris, Archibald, Ribeiro, Pedregosa, {van Mulbregt}, and {SciPy 1.0 Contributors}]{2020SciPy-NMeth}
Virtanen,~P. \latin{et~al.}  {{SciPy} 1.0: Fundamental Algorithms for Scientific Computing in Python}. \emph{Nat. Methods} \textbf{2020}, \emph{17}, 261--272\relax
\mciteBstWouldAddEndPuncttrue
\mciteSetBstMidEndSepPunct{\mcitedefaultmidpunct}
{\mcitedefaultendpunct}{\mcitedefaultseppunct}\relax
\EndOfBibitem
\bibitem[Broyden(1970)]{Broyden1970TheCO}
Broyden,~C.~G. The Convergence of a Class of Double-rank Minimization Algorithms 1. General Considerations. \emph{IMA J. Appl. Math.} \textbf{1970}, \emph{6}, 76--90\relax
\mciteBstWouldAddEndPuncttrue
\mciteSetBstMidEndSepPunct{\mcitedefaultmidpunct}
{\mcitedefaultendpunct}{\mcitedefaultseppunct}\relax
\EndOfBibitem
\bibitem[Fletcher(1970)]{Fletcher1970}
Fletcher,~R. A new approach to variable metric algorithms. \emph{Comput. J.} \textbf{1970}, \emph{13}, 317--322\relax
\mciteBstWouldAddEndPuncttrue
\mciteSetBstMidEndSepPunct{\mcitedefaultmidpunct}
{\mcitedefaultendpunct}{\mcitedefaultseppunct}\relax
\EndOfBibitem
\bibitem[Goldfarb(1970)]{Goldfarb1970}
Goldfarb,~D. A Family of Variable-Metric Methods Derived by Variational Means. \emph{Math. Comput.} \textbf{1970}, \emph{24}, 23--26\relax
\mciteBstWouldAddEndPuncttrue
\mciteSetBstMidEndSepPunct{\mcitedefaultmidpunct}
{\mcitedefaultendpunct}{\mcitedefaultseppunct}\relax
\EndOfBibitem
\bibitem[Shanno(1970)]{Shanno1970}
Shanno,~D.~F. Conditioning of Quasi-Newton Methods for Function Minimization. \emph{Math. Comput.} \textbf{1970}, \emph{24}, 647--656\relax
\mciteBstWouldAddEndPuncttrue
\mciteSetBstMidEndSepPunct{\mcitedefaultmidpunct}
{\mcitedefaultendpunct}{\mcitedefaultseppunct}\relax
\EndOfBibitem
\bibitem[Mishra and Ram(2019)Mishra, and Ram]{bfgs2019}
Mishra,~S.~K.; Ram,~B. \emph{Introduction to Unconstrained Optimization with R}; Springer Singapore: Singapore, 2019; pp 245--289\relax
\mciteBstWouldAddEndPuncttrue
\mciteSetBstMidEndSepPunct{\mcitedefaultmidpunct}
{\mcitedefaultendpunct}{\mcitedefaultseppunct}\relax
\EndOfBibitem
\bibitem[Witek \latin{et~al.}(2002)Witek, Choe, Finley, and Hirao]{isa-witek}
Witek,~H.~A.; Choe,~Y.-K.; Finley,~J.~P.; Hirao,~K. Intruder state avoidance multireference M{\o}ller--Plesset perturbation theory. \emph{J. Comput. Chem.} \textbf{2002}, \emph{23}, 957--965\relax
\mciteBstWouldAddEndPuncttrue
\mciteSetBstMidEndSepPunct{\mcitedefaultmidpunct}
{\mcitedefaultendpunct}{\mcitedefaultseppunct}\relax
\EndOfBibitem
\bibitem[Nijegorodov \latin{et~al.}(1997)Nijegorodov, Ramachandran, and Winkoun]{nijegorodov1997}
Nijegorodov,~N.; Ramachandran,~V.; Winkoun,~D. The dependence of the absorption and fluorescence parameters, the intersystem crossing and internal conversion rate constants on the number of rings in polyacene molecules. \emph{Spectrochim. Acta A} \textbf{1997}, \emph{53}, 1813--1824\relax
\mciteBstWouldAddEndPuncttrue
\mciteSetBstMidEndSepPunct{\mcitedefaultmidpunct}
{\mcitedefaultendpunct}{\mcitedefaultseppunct}\relax
\EndOfBibitem
\bibitem[Nakagawa \latin{et~al.}(2024)Nakagawa, Kamoshita, Mizukami, Sudo, and Ohnishi]{Nakagawa2024}
Nakagawa,~Y.~O.; Kamoshita,~M.; Mizukami,~W.; Sudo,~S.; Ohnishi,~Y.-y. ADAPT-QSCI: Adaptive construction of an input state for quantum-selected configuration iteraction. \emph{J. Chem. Theory Comput.} \textbf{2024}, \emph{20}, 10817--10825\relax
\mciteBstWouldAddEndPuncttrue
\mciteSetBstMidEndSepPunct{\mcitedefaultmidpunct}
{\mcitedefaultendpunct}{\mcitedefaultseppunct}\relax
\EndOfBibitem
\bibitem[Sugisaki \latin{et~al.}(2024)Sugisaki, Kanno, Itoko, Sakuma, and Yamamoto]{sugisaki2024}
Sugisaki,~K.; Kanno,~S.; Itoko,~T.; Sakuma,~R.; Yamamoto,~N. Hamiltonian simulation-based quantum-selected configuration interaction for large-scale electronic structure calculations with a quantum computer. 2024; \url{https://arxiv.org/abs/2412.07218}\relax
\mciteBstWouldAddEndPuncttrue
\mciteSetBstMidEndSepPunct{\mcitedefaultmidpunct}
{\mcitedefaultendpunct}{\mcitedefaultseppunct}\relax
\EndOfBibitem
\bibitem[Mikkelsen and Nakagawa(2025)Mikkelsen, and Nakagawa]{mikkelsen2025}
Mikkelsen,~M.; Nakagawa,~Y.~O. Quantum-selected configuration interaction with time-evolved state. 2025; \url{https://arxiv.org/abs/2412.13839}\relax
\mciteBstWouldAddEndPuncttrue
\mciteSetBstMidEndSepPunct{\mcitedefaultmidpunct}
{\mcitedefaultendpunct}{\mcitedefaultseppunct}\relax
\EndOfBibitem
\end{mcitethebibliography}
\end{document}